%% file: main.tex
\RequirePackage{etex}
\documentclass[12pt]{article}
\input{preamble}

\title{\fontsize{18}{22}\selectfont Breaking the Square-Root Barrier for Linear Swap Regret in Polyhedral Games}
\author[1,2]{Vasilis Pollatos\thanks{\texttt{vaspoll@math.uoa.gr}}}
\author[3]{Alberto Marchesi\thanks{\texttt{alberto.marchesi@polimi.it}}}
\affil[1]{Department of Mathematics, National and Kapodistrian University of Athens}
\affil[2]{Archimedes, Athena Research Center, Greece}
\affil[3]{Politecnico di Milano}
\date{}
\begin{document}
\maketitle

\input{sections/abstract}

\input{sections/introduction}
\input{sections/related-work}

\input{sections/preliminaries}

\input{sections/algorithmic-framework}

\input{sections/regret-analysis}

\input{sections/conclusion}

\section*{Acknowledgments}

Alberto Marchesi is supported by the contribution of Ministero dell’Università e della ricerca pursuant to D.D. n. 18010 of 12 November 2025 – BANDO FIS 3. Project FIS-2024-05736 (Starting Grant), title: “Towards a trustworthy strategic use of data in machine learning pipelines” (STRATDATA). CUP: D53C25002380001.
Vasilis Pollatos acknowledges financial support for his doctoral research
from the Archimedes Research Unit at the Athena Research Center.

\bibliographystyle{manuscript}
\bibliography{references}
\clearpage
\appendix
\input{appendices/related-work}

\input{appendices/polyhedral-geometry}

\input{appendices/treeplex-representation}

\input{appendices/barrier-comparison}

\input{appendices/reset-stability}

\input{appendices/resolvent-trace}

\input{appendices/fixed-point-stability}

\input{appendices/self-play-regret}

\input{appendices/finite-precision}

\input{appendices/unknown-horizons}

\input{appendices/utc-implementation}

\input{appendices/experiments}

\end{document}

%% file: preamble.tex
\usepackage[margin=.95in]{geometry}
\usepackage[T1]{fontenc}
\usepackage[utf8]{inputenc}
\usepackage{lmodern}
\usepackage[english]{babel}
\usepackage{fullpage}
\usepackage{authblk}
\usepackage{amsmath,amssymb,amsfonts,amsthm,mathtools}
\usepackage{bm,booktabs,tabularx,array,microtype,graphicx}
\usepackage[table]{xcolor}
\usepackage{tcolorbox}
\tcbuselibrary{breakable,skins}
\definecolor{TheoremBlueBack}{RGB}{245,249,255}
\definecolor{LemmaGreenBack}{RGB}{247,252,247}
\usepackage{enumitem,nicefrac,url,footmisc}
\usepackage{algorithm,algpseudocode,float}
\usepackage{natbib}
\setcitestyle{authoryear,round,citesep={;},aysep={,},yysep={;}}
\usepackage[hidelinks,colorlinks=true,linkcolor=red,citecolor=red]{hyperref}
\definecolor{CardinalRed}{HTML}{C41E3A}
\definecolor{Dartmouth}{HTML}{00693E}
\definecolor{Purple}{RGB}{146,42,158}
\colorlet{MyRed}{CardinalRed}
\colorlet{MyGreen}{Dartmouth}
\hypersetup{
    pdftitle={Breaking the Square-Root Barrier for Linear Swap Regret in Polyhedral Games},
    pdfauthor={Vasilis Pollatos and Alberto Marchesi},
    pdfkeywords={},
    pdfborder=0 0 0,
    pdfpagemode=UseNone,
    colorlinks=true,
    linkcolor=MyRed,
    citecolor=MyGreen,
    filecolor=Purple,
    urlcolor=blue,
}
\allowdisplaybreaks
\newtheorem{theorem}{Theorem}
\newtheorem{lemma}[theorem]{Lemma}
\newtheorem{proposition}[theorem]{Proposition}
\newtheorem{corollary}[theorem]{Corollary}

\theoremstyle{definition}
\tcbset{paperstatement/.style={enhanced,breakable,frame hidden,boxrule=0pt,
  colback=TheoremBlueBack,
  top=2pt,left=2pt,right=2pt,bottom=2pt}}
\tcolorboxenvironment{theorem}{paperstatement}
\tcolorboxenvironment{lemma}{paperstatement,colback=LemmaGreenBack}
\tcolorboxenvironment{proposition}{paperstatement,colback=LemmaGreenBack}
\tcolorboxenvironment{corollary}{paperstatement}
\tcolorboxenvironment{remark}{paperstatement,colback=LemmaGreenBack}
\tcolorboxenvironment{definition}{paperstatement,colback=LemmaGreenBack}
\newcommand{\R}{\mathbb R}
\newcommand{\E}{\mathbb E}
\newcommand{\ri}{\operatorname{ri}}
\newcommand{\aff}{\operatorname{aff}}
\newcommand{\LinReg}{\operatorname{LinReg}}
\newcommand{\TV}{\operatorname{TV}}

\newcommand{\poly}{\operatorname{poly}}
\newcommand{\KL}{D_{\mathrm{KL}}}
\newcommand{\norm}[1]{\left\lVert #1\right\rVert}
\newcommand{\ip}[2]{\left\langle #1,#2\right\rangle}
\newcommand{\eps}{\varepsilon}
\newcommand{\cZ}{\mathcal Z}
\newcommand{\tr}{\operatorname{tr}}
\newcommand{\adj}{\operatorname{adj}}

\graphicspath{{figures/}}

%% file: sections/abstract.tex
\begin{abstract}
We study linear swap regret in self-play against all affine self-maps of each player's strategy polytope. Efficient adversarial bounds for this class scale as $\sqrt T$. On compact polytopes given by linear inequalities, our deterministic uncoupled dynamics achieve $\widetilde{\mathcal{O}}(\poly(D)T^{1/3})$ cumulative regret, where $D$ is the binary length of the game description. The bound is independent of the number of vertices. In perfect-recall $n$-player extensive-form games with at most $M$ terminal sequences per player, we obtain $\mathcal{O}(n^2M^4\log T)$ regret against all linear deviations, extending the logarithmic trigger-regret bound of \citet{anagnostides2023}. This class can strictly contain the trigger deviations \citep{farina2023} and consists exactly of the deviations implementable by untimed communication (UTC), in which a player interleaves moves with mediator queries along a separate feasible path \citep{zhang2024}. The proof combines optimistic log-barrier learning on a compact matrix lift with stability of the Gordon fixed point, using treeplex geometry in extensive form.
\end{abstract}

%% file: sections/introduction.tex
\section{Introduction}\label{sec:intro}
No-regret learning in games becomes substantially more challenging as the class of comparators grows more expressive. External regret compares the learner's performance against that of a fixed action and leads to coarse correlated equilibria, whereas swap regret allows for action-dependent substitutions and leads to correlated equilibria \citep{fostervohra1997,hartmascolell2000,blummansour2007}. The $\Phi$-regret framework \citep{gordon2008} generalizes these notions to arbitrary classes of transformations. On a convex strategy set, \emph{linear deviations} are its affine self-maps. They define linear swap regret (also called linear $\Phi$-regret) and linear correlated equilibrium \citep{farina2023,daskalakis2025}.

The computational challenge is particularly pronounced in games whose strategy spaces admit compact polyhedral representations but have exponentially many vertices. In explicitly represented finite games, bounds on full swap regret in self-play have improved from logarithmic \citep{anagnostides2022swap} to sublogarithmic \citep{tsuchiya2026} and eventually horizon-independent rates \citep{mai2026constant}. However, these bounds still exhibit polynomial dependence on the number of explicitly represented actions. Perfect-recall extensive-form games (EFGs) provide a prominent setting where this distinction is crucial: the sequence-form strategy space admits a polynomial-size representation, whereas the normal form contains exponentially many contingent plans \citep{koller1996}. Hence fast full-swap algorithms can take exponential time in the game-tree size. Over $T$ rounds of play, efficient adversarial regret bounds for linear
deviations scale as $\sqrt{T}$ on both general convex sets
\citep{daskalakis2025,anagnostides2026} and the sequence-form polytopes
of EFGs \citep{farina2023}. 

\citet{farina2023} identify why the optimistic analysis based on regret bounded by variation in utilities (RVU), which underlies the $o(\sqrt{T})$ bounds on cumulative trigger regret in self-play, does not directly extend to linear deviations. The logarithmic-regret construction of \citet{anagnostides2023} exploits the fact that trigger fixed points admit rational representations with positive coefficients, inherited from the stationary distributions of local Markov chains. In contrast, the Gordon fixed point of a general linear deviation has no known analogous local characterization. The UTC formulation of \citet{zhang2024} provides a polynomial-size domain for optimizing over all linear deviations. However, exploiting the negative movement term in the optimistic regret bound requires controlling how changes in the deviation map affect the corresponding fixed point played by the learner. This leads to the following question:

\begingroup
\def\researchquestion#1{%
  \par\addvspace{.45\baselineskip}%
  {\centering\itshape
   \leftskip=.03\linewidth plus .25\linewidth\relax
   \rightskip=\leftskip
   \interlinepenalty=10000
   \hyphenpenalty=10000
   #1\par}%
  \addvspace{.45\baselineskip}%
}

\researchquestion{\textbf{\textup{Q1:}}~Can efficient uncoupled dynamics in self-play achieve
$o(\sqrt{T})$ linear $\Phi$-regret on polytopes with succinct $H$-representations,
with polynomial dependence on the input size?}

\enlargethispage{\baselineskip}
In perfect-recall EFGs, \citet{anagnostides2023} attain logarithmic trigger
regret in self-play. A trigger switches to a fixed continuation on a specified
recommendation at an information set. Thus empirical play is an
$\mathcal{O}(\log T/T)$-approximate extensive-form correlated equilibrium (EFCE).
UTC deviations also allow mediator queries along a feasible hypothetical path before
acting. UTC regret equals linear $\Phi$-regret \citep[Thm.~4.1]{zhang2024}.

In the signaling EFCE of \citet{vonstengelforges2008},
\citet[Example~E.1]{farina2023} exhibit a profitable linear deviation.
To connect this equilibrium example to no-regret learning, consider repeated play that cycles through the four equally likely joint plans. After each complete cycle, the empirical distribution of play coincides with the equilibrium distribution. In the underlying signaling game, a student privately observes their type (good or bad) and sends one of two signals to a professor, who then decides whether to accept the student. The professor accepts exactly the signal recommended to the good type,
so the good type already secures acceptance by obeying. A trigger
deviation can use only the recommendation for the student's actual type.
Each bad-type recommendation occurs in two equally likely plans. In one,
the professor accepts the recommended signal. In the other, the professor
accepts the opposite signal. Thus obeying and switching each yield
acceptance with probability~$1/2$, so neither type gains from a trigger
deviation. Under UTC, even a student of the bad type can ask the mediator
which signal it would recommend to the good type. Sending that signal
guarantees acceptance. Over complete cycles, the student's trigger regret
is zero, while this UTC deviation raises the student's expected cumulative
payoff from $7T/2$ to $5T$, a gain of $3T/2$.

Motivated by the logarithmic trigger regret achieved by optimistic dynamics
\citep{anagnostides2023} and by the ability of UTC deviations to profit where
no trigger can, we turn to our second question, also posed by
\citet[Sec.~5]{farina2023} and \citet[Question~3]{zhang2024}:

\researchquestion{\textbf{\textup{Q2:}}~Can efficient uncoupled dynamics for perfect-recall EFGs
achieve $\mathcal{O}(\log T)$ linear~$\Phi$-regret in self-play, with polynomial
dependence on the size of the game tree?}
\enlargethispage{\baselineskip}
\endgroup

We answer both questions Q1 and Q2 in the affirmative. Let $D$ denote the binary encoding length of the compact game description. We represent every linear deviation by a nonnegative matrix in a compact polytope. The optimistic log-barrier update keeps the learner’s matrices strictly positive and controls how much they change between rounds. For arbitrary inequality-described ($H$-)polytopes, a weak reset stabilizes the Gordon fixed point and gives a $T^{1/3}$ rate; for EFGs, recursive treeplex geometry stabilizes the \emph{unreset} fixed point and yields logarithmic regret against every linear deviation.

\begin{theorem}[Main results]\label{thm:informal}
We consider normalized $n$-player multi-affine games with compact rational strategy polytopes and polynomial-time full-information feedback.  For perfect-recall EFGs, $M$ denotes the maximum number of terminal sequences of any player.  Algorithm~\ref{alg:exact} admits deterministic uncoupled implementations such that, for every player,
\[
\LinReg_i(T)=
\begin{cases}
\widetilde{\mathcal{O}}(\poly(D)T^{1/3}),&\text{arbitrary compact $H$-polytopes},\\
\mathcal{O}(n^2M^4\log(T+1)),&\text{{perfect-recall EFGs}},
\end{cases}
\]
The bounds are independent of the number of vertices, and each round uses $\poly(D,\log(T+1))$ bit operations.
\end{theorem}

\paragraph{Technical overview.}
Our algorithmic framework, which is formally defined in Algorithm~\ref{alg:exact},  couples two geometries: optimistic follow-the-regularized-leader (FTRL) moves in a compact positive lift, while Gordon's reduction converts each deviation into the fixed-point strategy actually played.  The proof must therefore transfer slow \emph{deviation movement} into slow \emph{strategy movement}.  The optimistic FTRL regret bound subtracts two Bregman divergences per round, one for the prediction step and one for the correction step. Smoothness of the multi-affine payoffs controls the cumulative squared prediction error in terms of the squared path length of play. Fixed-point stability then bounds this path length by a multiple of the sum of those same divergences. Theorem~\ref{thm:exact} isolates this cancellation, leaving only the geometry-specific stability bound.

For arbitrary $H$-polytopes, exact fixed points can be unstable.  A weak reset makes the Gordon map contractive, but direct perturbation would square the inverse reset strength.  We instead represent the reset fixed point as the mean endpoint of a regenerative random composition.  The KL chain rule charges only the expected trajectory length, so an entropic-to-log-barrier comparison yields a single inverse-reset factor.  Balancing this stability cost with reset bias gives the $T^{1/3}$ rate.

Perfect-recall EFGs admit a sharper route.  Their realization-plan polytopes restricted to terminal sequences are treeplexes built recursively from products and joins, inducing a sum/max norm in which every feasible deviation matrix is nonexpansive.  A resolvent-trace sign survives this recursion.  Applied to a regularized Perron equation, that sign makes the scalar denominator in the adjugate formula monotone as the two coefficients of a feasible matrix family vary; a one-dimensional Bernstein inequality then converts denominator control into a polynomial derivative bound for the fixed point.  This removes the reset entirely and yields logarithmic regret for the full linear/UTC class.

\begin{table}[!t]
\centering
\footnotesize
\setlength{\tabcolsep}{4pt}
\renewcommand{\arraystretch}{1.12}
\hypersetup{citecolor=blue}
\caption{Cumulative regret over $T\ge2$ rounds in normalized $n$-player
games. Here $N\ge2$ is the largest number of pure strategies of any
player in the normal form. In polyhedral games, these strategies may be represented implicitly and $N$ can be exponential in $d$, the largest
intrinsic dimension of a player's strategy polytope. Throughout the
table, $D$ is the binary length of the compact game description and
$\poly(\cdot)$ denotes polynomial factors. The proximal bound fixes
Lipschitz and smoothness constants and strategy-set diameters.}
\label{tab:comparison}
\begin{tabularx}{\textwidth}{@{}>{\raggedright\arraybackslash}p{.27\textwidth}>{\raggedright\arraybackslash}p{.25\textwidth}>{\raggedright\arraybackslash}p{.17\textwidth}>{\raggedright\arraybackslash}X@{}}
\toprule
Reference & Setting / representation & Comparator & Cumulative regret\\
\midrule
\citet{mai2026constant} & Normal form; self-play & Full swap & $\mathcal{O}(\poly(n,N))$\\
\rowcolor{black!6}
\citet{dagan2024swap};\newline\citet{pengrubinstein2024swap} & Normal form / EFG pure plans; adversarial & Full swap & $\mathcal{O}(T\log\log(NT)/\log T)$\\
\citet{anagnostides2023} & Perfect-recall EFGs; self-play & Trigger & $\mathcal{O}(\poly(D)\log T)$\\
\rowcolor{black!6}
\citet{cai2026proximal} & Smooth convex games; self-play & Proximal & $\mathcal{O}(\poly(n)T^{1/4})$\\
\citet{farina2023,zhang2024} & Perfect-recall EFGs; adversarial & Linear $\Phi$ / UTC & $\mathcal{O}(\poly(D)T^{1/2})$\\
\rowcolor{black!6}
\citet{anagnostides2026} & John-preconditioned convex sets; adversarial & Linear $\Phi$ & $\mathcal{O}(\poly(d)T^{1/2})$\\
\textbf{This work} & Compact $H$-polytopes; self-play & Linear $\Phi$ & $\widetilde{\mathcal{O}}(\poly(D)T^{1/3})$\\
\rowcolor{black!6}
\textbf{This work} & Perfect-recall EFGs; self-play & Linear $\Phi$ / UTC & $\mathcal{O}(\poly(D)\log T)$\\
\bottomrule
\end{tabularx}
\end{table}

%% file: sections/related-work.tex
\section{Further Related Work}\label{sec:related}
\paragraph{Fast swap regret in normal form.}
Earlier self-play guarantees in normal-form games include the $T^{1/4}$ bound of \citet{chenpeng2020} and the polylogarithmic bound of \citet{anagnostides2022near}.  Scale-invariant dynamics attain logarithmic swap regret without prior knowledge of utility scales \citep{tsuchiya2026scale}.  A horizon-independent guarantee is also available for games in which the sum of the players' external regrets is always nonnegative \citep{hait2025adaptive}.  Appendix~\ref{app:related} discusses this restriction and the distinction from recent constant external-regret bounds.

\paragraph{Swap regret with large action spaces.}
The \emph{Swap Regret 2.0} reduction of \citet{dagan2024swap} and the concurrent work of \citet{pengrubinstein2024swap} give cumulative full swap regret $\mathcal{O}(T\log\log(NT)/\log T)$ against adversarial utilities, improving the dependence on the implicit action count $N$, which may be exponential in the polyhedral dimension. For combinatorial bandits on a ground set of size $k$,
\citet[Eq.~(2) and Thm.~4.1]{kontogiannis2026swap} adapt this multiscale approach
to aggregate reward feedback and obtain
$\mathcal{O}(T\log(k\log T)/\log T)$ swap pseudo-regret against an
oblivious adversary. Even in EFGs, attaining inverse-polynomial average full swap regret can require superpolynomially many rounds against an oblivious adversary \citep{daskalakis2024swap}. For multilinear games on explicit polytopes, \citet[Thm.~4]{anagnostides2026ce} show that for any prescribed polynomial support bound, computing a CE at a suitable inverse-polynomial accuracy is PPAD-hard under the PCP-for-PPAD conjecture.  For finite-horizon Markov games with a fixed number of players, \citet{anagnostides2026markov} give an FPTAS for normal-form correlated equilibria.

\paragraph{Other comparator classes and compact learning.}
Randomized weighted majority and kernelized multiplicative weights avoid pure-action enumeration in several structured games \citep{beaglehole2023,farina2022kernel,kontogiannis2025kernel}, but their equilibrium guarantees are based on external regret.  In Lipschitz, smooth convex games, \citet[Thm.~4]{cai2026proximal} obtain a $T^{1/4}$ self-play rate against proximal maps of convex functions. Their affine comparison covers deviations with a symmetric linear part.

%% file: sections/preliminaries.tex
\section{Preliminaries}\label{sec:model}
\paragraph{Linear algebra and divergences.}
For a convex set $K$, $\aff K$ and $\ri K$ denote its affine hull and relative interior. For a square matrix $A$, $\tr A$ and $\adj A$ denote its trace and adjugate. For a differentiable convex function $F$, we write $D_F(v,u)=F(v)-F(u)-\langle\nabla F(u),v-u\rangle$ for its Bregman divergence. For probability laws $P,Q$, $\KL(P\|Q)$ and $\TV(P,Q)$ denote KL divergence and total variation.

\paragraph{Polyhedral games.}
A \emph{polyhedral game} consists of a player set $[n]:=\{1,\ldots,n\}$, a nonempty compact strategy polytope $Q_i$ for each player $i$, and a scalar loss $L_i:\prod_{j=1}^n Q_j\to[-1,1]$ for each $i$. Each $L_i$ is affine in each player's strategy separately. This includes the $0/1$-polyhedral games of \citet{farina2022kernel,kontogiannis2025kernel}. We write $d_i:=\dim(\aff Q_i)$. At a fixed opponents' profile $x_{-i}$, let $\ell_i(x_{-i})$ denote the linear part of player $i$'s loss in its own strategy. Thus, for $u,v\in Q_i$,
\[
L_i(u,x_{-i})-L_i(v,x_{-i})
=\langle\ell_i(x_{-i}),u-v\rangle.
\]
This defines $\ell_i(x_{-i})$ on $\operatorname{span}(Q_i-Q_i)$.

We measure the distance between strategies by the largest difference in loss that an affine function taking values in $[0,1]$ can assign to them. For $u,v\in Q_i$, set
\[
\|u-v\|_i
:=\sup\bigl\{|f(u)-f(v)|:\ f:Q_i\to[0,1]\text{ affine}\bigr\}.
\]
Thus $\|u-v\|_i\le1$ for feasible strategies $u,v$. On a probability simplex this is the total variation distance $\tfrac12\|u-v\|_1$. In general it extends by homogeneity to a norm on $\operatorname{span}(Q_i-Q_i)$, as shown in Appendix~\ref{app:polyhedral}. Its dual measures the largest loss difference between feasible strategies, $\|g\|_{i,*}=\max_{u,v\in Q_i}|\langle g,u-v\rangle|$. Thus the normalization $L_i\in[-1,1]$ bounds unilateral loss differences by two. For every affine self-map $\phi:Q_i\to Q_i$, composing an affine test
$f:Q_i\to[0,1]$ with $\phi$ gives another such test. Hence $\phi$ is
nonexpansive in the strategy norm:
$\|\phi(u)-\phi(v)\|_i\le\|u-v\|_i$ for all $u,v\in Q_i$.
Multi-affinity gives
\begin{equation}
\|\ell_i(x_{-i})\|_{i,*}\le2,\qquad
\|\ell_i(x_{-i})-\ell_i(y_{-i})\|_{i,*}^2
\le16(n-1)\sum_{j\ne i}\|x_j-y_j\|_j^2.
\label{eq:smoothness}
\end{equation}

\paragraph{Online learning and linear $\Phi$-regret.}
We study repeated simultaneous play with full-information feedback \citep{farina2022kernel,kontogiannis2025kernel}. At round $t$, each player $i$ selects $x_i^t\in Q_i$ using its own past strategies and loss vectors. Once the players have selected the profile $x^t=(x_1^t,\ldots,x_n^t)$, player $i$ incurs $L_i(x_i^t,x_{-i}^t)$ and observes $\ell_i^t:=\ell_i(x_{-i}^t)$. This feedback specifies the loss difference between any two of its strategies. Thus each player faces an online linear optimization problem. The dynamics are \emph{uncoupled}. Apart from common size and horizon parameters, updates use only the player's own strategy polytope and feedback history. Our guarantees concern \emph{self-play}, where all players follow the prescribed learning rule.

Player $i$'s linear deviations are the affine self-maps $\Phi_i=\{\phi:Q_i\to Q_i:\phi\text{ affine}\}$. An affine map $x\mapsto Bx+b$ is linear in the augmented coordinates $(x,1)$. We use the loss convention of \citet[Eq.~(1)]{daskalakis2025}. Player $i$'s cumulative linear $\Phi$-regret is
\[
\LinReg_i(T):=\sum_{t=1}^T L_i(x^t)
  -\min_{\phi\in\Phi_i}\sum_{t=1}^T L_i\bigl(\phi(x_i^t),x_{-i}^t\bigr).
\]
The minimum chooses a single linear deviation in hindsight for all $T$ rounds, holding the opponents' strategies fixed at their observed values. A distribution on strategy profiles is an $\varepsilon$-\emph{linear correlated equilibrium} if each player's expected loss reduction from any such deviation is at most $\varepsilon$. Under the empirical distribution, which assigns weight $1/T$ to each round, the largest such reduction for player $i$ is exactly $\LinReg_i(T)/T$.

\paragraph{Extensive-form games.}
We use the standard finite perfect-recall model; see \citet{koller1996,farina2023} for detailed definitions and the sequence-form representation. Write $\mathcal J_i$ for player $i$'s information sets, $\mathcal A_j$ for the actions at $j$, and $p(j)$ for the player's sequence immediately before $j$. The sequence set is $\Sigma_i:=\{\varnothing\}\cup\{ja:j\in\mathcal J_i,\ a\in\mathcal A_j\}$, where $\varnothing$ is the empty sequence and $ja$ records taking action $a$ at $j$. We use $\Sigma_i^\perp:=\{\sigma\in\Sigma_i:p(j)\ne\sigma\text{ for every }j\in\mathcal J_i\}$ for the terminal sequences of this player. A play history can continue after the player's last decision. We set $m_i:=|\Sigma_i^\perp|$ and $M:=\max_i m_i$. The realization-plan polytope is
\begin{equation}
Q_i=\left\{x\ge0:x_\varnothing=1,\ \sum_{a\in\mathcal A_j}x_{ja}=x_{p(j)}\ (j\in\mathcal J_i)\right\}.
\label{eq:sequence}
\end{equation}
Here $x_\sigma$ is the product of the player's own action probabilities along $\sigma$. Since $x_\varnothing=1$, linear deviations on $Q_i$ can be written as linear maps of realization plans. A realization plan is determined by its coordinates on $\Sigma_i^\perp$. Restricting to those coordinates gives an affinely isomorphic treeplex and preserves all linear deviations. In an untimed communication (UTC) deviation the player can query a mediator for recommended actions without matching the timing of play in the game. These deviations induce exactly the linear deviations on realization plans \citep[Thm.~4.1]{zhang2024}. Appendix~\ref{app:efg-prelims} gives the recursive construction and
shows how to recover the full realization plan from its terminal-sequence
coordinates.

%% file: sections/algorithmic-framework.tex
\section{Algorithmic Framework}\label{sec:algorithm}
\paragraph{Compact affine lift.}
We represent strategies in coordinates in which deviations can be learned as nonnegative matrices. For a general polytope, the coordinates are the slacks of the inequalities that are not tight everywhere. For an EFG, they are the realization weights on the player's terminal sequences. 
In either case, the change of coordinates is an affine bijection from
$Q_i$ onto a polytope $P_i\subseteq\R_+^{m_i}$. Here $m_i$ counts the
slack coordinates for a general polytope or the terminal sequences for
an EFG. Choose a rational reference point $s_i^0\in P_i$ with all
coordinates strictly positive. The differences $s-s_i^0$ between
feasible points and this reference point span a linear space $W_i$ of
dimension $d_i=\dim Q_i$. The affine equations of the representation
restrict $s$ to $s_i^0+W_i$, and imposing nonnegativity gives
\[
P_i=(s_i^0+W_i)\cap\R_+^{m_i}.
\]

As shown in Appendix~\ref{app:polyhedral}, we can choose positive weights
$w_i$ so that $w_i^\top s=1$ for every $s\in P_i$. This normalization
allows us to express affine functions in linear form: on $P_i$, the affine
function $g^\top s+b$ equals $(g+bw_i)^\top s$.

The change of coordinates also preserves the strategy distance defined
in Section~\ref{sec:model}. Every affine loss $Q_i\to[0,1]$ corresponds
to an affine loss $P_i\to[0,1]$ with the same values on corresponding
strategies, and conversely. Thus the largest loss difference between
two strategies is the same in either representation.

We continue to write $x_i$ for the strategy in these coordinates and choose an ambient vector $\ell_i\in\R^{m_i}$ representing the original unilateral loss differences on $P_i$. Any two such choices give losses differing by a constant on $P_i$, so they give the same deviation comparisons.

A nonnegative matrix that preserves the affine slice $s_i^0+W_i$ maps $P_i$ into itself. We use these matrices as the learner's decision set and the entrywise log barrier on its positive part. With $A>0$ denoting entrywise strict positivity, define
\begin{equation}
\cZ_i=\{A\in\R_+^{m_i\times m_i}:
A(s_i^0+W_i)\subseteq s_i^0+W_i\},\qquad
R_i(A)=-\sum_{a,b=1}^{m_i}\log A_{ab}.
\label{eq:slack-lift}
\end{equation}

\begin{proposition}[Compact representation of linear deviations]\label{prop:slack}
For each player $i$, the lift $\cZ_i$ has the following three properties.
\begin{enumerate}[label=\textup{(\roman*)},nosep,leftmargin=*]
\item $\cZ_i$ is a compact polytope defined by linear equalities and nonnegativity. It contains the strictly positive matrix $s_i^0 w_i^\top$.
\item The restrictions $s\mapsto As$ for $A\in\cZ_i$ are exactly the affine self-maps of $P_i$, and hence represent every linear deviation on $Q_i$.
\item Every $A\in\cZ_i$ with $A>0$ has a unique fixed point $s_A\in P_i$. It is strictly positive and satisfies $w_i^\top s_A=1$.
\end{enumerate}
\end{proposition}

\citet{daskalakis2025} give an efficient membership test for affine maps when the target polytope is described by inequalities and linear optimization over the source is efficient. Every $A\in\cZ_i$ maps $P_i$ into itself and thus induces an
affine self-map of $Q_i$ through the chosen affine isomorphism.
Conversely, row-wise LP duality shows that every affine self-map
of $Q_i$ is induced by some $A\in\cZ_i$. Each output coordinate is represented by a nonnegative combination of input coordinates on $P_i$, using $w_i^\top s=1$ to absorb constants. Compactness and Perron--Frobenius theory yield the fixed-point statement (Appendix~\ref{app:polyhedral}).  The lift has two roles: it represents the full hindsight comparator class in polynomial size, and its positive matrices provide the fixed points required by Gordon's reduction. The learner updates in matrix space, where $R_i$ controls movement; the analysis below transfers this control to the strategies played.

We set $p_i:=d_i(d_i+1)$, the number of affine-map coefficients in intrinsic coordinates, and $p:=\max_i p_i$. We omit singleton strategy sets from the updates since their regret is zero. We use $M=\max_i m_i$ also for general polytopes.  The general-polytope mode uses $\lambda>0$: for a fixed reset point $q_i^0\in P_i$ (for example, the rational anchor $s_i^0$), $x_A^\lambda$ is the unique solution of $x=(1-\lambda)Ax+\lambda q_i^0$, and the reset makes the map contractive.  EFG mode sets $\lambda=0$ and plays $x_A^0=s_A$, the unique positive solution of $As_A=s_A$ normalized by $w_i^\top s_A=1$. Section~\ref{sec:efg-stability} proves stability using treeplex geometry.

\begin{algorithm}[!htp]
\caption{ORBITS for player $i$ (optimistic FTRL with Gordon's reduction)}\label{alg:exact}
\begin{algorithmic}[1]
\Require $\cZ_i$, reset $q_i^0$, parameters $\lambda\in[0,1)$, $\gamma>0$
\State Initialize the deviation loss $c_i^0\gets0$
\For{$t=1,2,\ldots$}
\State $A_i^t\gets\displaystyle\arg\min_{A\in\cZ_i,\,A>0}
    \Bigl\{\gamma R_i(A)+\sum_{\tau=1}^{t-1}c_i^\tau(A)+c_i^{t-1}(A)\Bigr\}$
\State Find $x_i^t\in P_i$ satisfying $x_i^t=(1-\lambda)A_i^t x_i^t+\lambda q_i^0$
\State Play $x_i^t$ and observe $\ell_i^t$
\State Set $c_i^t(A)\gets(1-\lambda)\langle\ell_i^t,Ax_i^t\rangle$ for $A\in\cZ_i$
\EndFor
\end{algorithmic}
\end{algorithm}

We call Algorithm~\ref{alg:exact} \emph{ORBITS} (Optimistic Regret via Barrier Iterates and Transformation Stability). The parameters $\lambda$ and $\gamma$ are chosen for general polytopes and EFGs in Sections~\ref{sec:general-polytopes} and~\ref{sec:efg-stability}, respectively.
We combine prediction, play, and feedback in one cycle.  The optimistic matrix update predicts the next deviation loss with the previous one (line~3), and Gordon's fixed-point step converts that matrix into the strategy actually played (line~4).  We choose the pulled-back cost (line~6) so that $c_i^t(A_i^t)-c_i^t(C)=\langle\ell_i^t,x_i^t-[(1-\lambda)Cx_i^t+\lambda q_i^0]\rangle$ for every comparator $C\in\cZ_i$.  Hence matrix regret is already the Gordon deviation comparison: only the explicit reset separates it from the original deviation, and that reset vanishes in EFG mode.  Movement and stability are therefore the remaining analytical bottleneck.

At the beginning of round $t$, the accumulated realized cost $\sum_{\tau=1}^{t-1}c_i^\tau$ summarizes the past, while $c_i^{t-1}$ is reused once as a prediction of the next deviation loss.  We therefore obtain a positive prediction matrix $A_i^t$ from the barrier-regularized minimization (line~3). The log barrier keeps each iterate $A_i^t$ strictly positive. Regret
is measured against all nonnegative matrices in $\cZ_i$, including
those with zero entries. By Proposition~\ref{prop:slack}, this covers
every linear deviation.

The matrix itself cannot be played, so we convert it into a feasible strategy through Gordon's fixed-point equation (line~4).  In general-polytope mode the reset makes the conversion contractive; in EFG mode $\lambda=0$ and the learner plays $s_A$ directly.  After the game reveals $\ell_i^t$, we pull the loss back to deviation space through $c_i^t(A)=(1-\lambda)\langle\ell_i^t,Ax_i^t\rangle$ and add that realized cost to the cumulative score (lines~5--6).  Optimism is therefore useful only when slow matrix movement survives the fixed-point conversion, which is precisely the stability question below.  One round (lines~3--6) requires one barrier-regularized convex optimization over the polynomial-size lift, one linear fixed-point solve, and linear feedback updates; Appendix~\ref{app:implementation} gives the deterministic finite-precision implementation and bit-complexity proof.

%% file: sections/regret-analysis.tex
\section{Regret Analysis}\label{sec:analysis}
The log barrier controls how far the selected deviation matrix moves between rounds. The variation in a player's losses, however, depends on the strategies played by its opponents. These strategies are fixed points of the selected maps, and a small change in a map can move its fixed point substantially. Section~\ref{sec:general-regret} first proves a regret bound conditional on fixed-point stability. Sections~\ref{sec:general-polytopes} and~\ref{sec:efg-stability} then prove the required stability with a reset for general polytopes and without a reset for EFGs.

\subsection{General regret bound}\label{sec:general-regret}
Optimistic FTRL controls movement of deviation matrices, while loss variation depends on movement of the strategies played. Fixed-point stability connects these two quantities. Let $A_{c,i}$ be the analytic center and $\bar A_i^t$ the correction leader formed from realized costs alone, without the extra prediction:
\[
A_{c,i}:=\arg\min_{A\in\cZ_i,\,A>0}R_i(A),\qquad
\bar A_i^t:=\arg\min_{A\in\cZ_i,\,A>0}\{\gamma R_i(A)+\textstyle\sum_{\tau=1}^{t-1}c_i^\tau(A)\}.
\]
We measure matrix movement and strategy movement by
\[
E_i:=\sum_{t=1}^T\!\bigl[D_{R_i}(A_i^t,\bar A_i^t)+D_{R_i}(\bar A_i^{t+1},A_i^t)\bigr],\qquad
V_i:=\sum_{t=2}^T\|x_i^t-x_i^{t-1}\|_i^2.
\]
The energy $E_i$ records the two directed Bregman steps between prediction and correction; $V_i$ is the squared path length of the actual strategies. We measure prediction error by the largest change since the previous round in the loss gap between two feasible deviations,
\begin{equation}
\Delta_i^t :=
 \max_{B,C\in\cZ_i}
 \big|(c_i^t(B)-c_i^t(C))-(c_i^{t-1}(B)-c_i^{t-1}(C))\big|.
\label{eq:prediction-gap}
\end{equation}
The standard optimistic-FTRL/RVU bound \citep[Thm.~19, Prop.~7]{syrgkanis2015}, in directed-Bregman form, gives, for every interior comparator $C$,
\begin{equation}
\sum_{t=1}^T[c_i^t(A_i^t)-c_i^t(C)]
\le \gamma D_{R_i}(C,A_{c,i})+\gamma^{-1}\sum_{t=1}^T(\Delta_i^t)^2-\tfrac\gamma2E_i.
\label{eq:rvu}
\end{equation}
The three terms are the comparator radius, the positive prediction error, and the negative barrier energy. We bound prediction error through the strategy path lengths $V_i$, then use fixed-point stability to bound each $V_i$ by $E_i$. This allows the two movement terms to cancel.

The next lemma relates the prediction error, the strategies played, and the movement of the deviation matrices.

\begin{lemma}[Self-play variation and fixed-point stability]\label{lem:variation-main}
\textup{(i) Prediction error.} In self-play,
\[
\sum_{i=1}^n\sum_{t=1}^T(\Delta_i^t)^2
\le 4n+32n^2\sum_iV_i.
\]
\textup{(ii) Strategy movement.} Suppose that, for player $i$ and all positive $A,B\in\cZ_i$, their played fixed points satisfy
$\|x_B^\lambda-x_A^\lambda\|_i^2\le C_*D_{R_i}(B,A)$, for some constant $C_*>0$. Then
\[
V_i\le 2C_*E_i.
\]
\end{lemma}
\begin{proof}[Proof sketch]
For $t\ge2$, the change in the loss gap for any two comparators $B,C\in\cZ_i$ is given by
\[
(c_i^t-c_i^{t-1})(B-C)=(1-\lambda)\bigl[\langle\ell_i^t-\ell_i^{t-1},(B-C)x_i^t\rangle+\langle\ell_i^{t-1},(B-C)(x_i^t-x_i^{t-1})\rangle\bigr].
\]
Equation~\eqref{eq:smoothness} controls the first term by opponent movement;nonexpansiveness of $B$ and $C$ in the strategy norm $\|\cdot\|_i$ and $\|\ell_i^{t-1}\|_{i,*}\le2$ bound the second by $4\|x_i^t-x_i^{t-1}\|_i$.  Squaring and summing gives the first inequality (the initialized round contributes $4n$). To compare consecutive plays, we apply fixed-point stability first
between $A_i^{t-1}$ and $\bar A_i^t$, then between $\bar A_i^t$ and
$A_i^t$. The triangle inequality and $(a+b)^2\le 2(a^2+b^2)$ then
give $V_i\le 2C_*E_i$ after summing over rounds.  Crucially, the two directed divergences generated by this bridge are exactly those already present with negative sign in~\eqref{eq:rvu}.  Appendix~\ref{app:closure} gives the full calculation.
\end{proof}

\begin{theorem}[Stability-to-regret]\label{thm:exact}
Suppose a common $C_*>0$ satisfies $\|x_B^\lambda-x_A^\lambda\|_i^2\le C_*D_{R_i}(B,A)$ for every player $i$ and all positive $A,B\in\cZ_i$. If $\gamma^2\ge128n^2C_*$, the exact iterates satisfy
\begin{equation}
\sum_i\LinReg_i(T)
\le\gamma\Bigl(\sum_i m_i^2\Bigr)\log(T+1)+\frac{4n}{\gamma}+2n(\lambda T+1).
\label{eq:master}
\end{equation}
The same right-hand side bounds $\LinReg_i(T)$ for each $i$, since the identity deviation saves zero loss and hence every player's regret is nonnegative.
\end{theorem}
\begin{proof}[Proof sketch]
After summing~\eqref{eq:rvu} over players, we apply
Lemma~\ref{lem:variation-main}(i) to bound the prediction error.
Under the theorem's fixed-point stability hypothesis,
part~(ii) then gives
\[
\frac{32n^2}{\gamma}\sum_i V_i
\le \frac{32n^2}{\gamma}\sum_i 2C_*E_i
=\frac{64n^2C_*}{\gamma}\sum_iE_i.
\]
The first step uses $V_i\le2C_*E_i$, and the second only collects constants.  Against the negative $-(\gamma/2)\sum_iE_i$ term, the net energy coefficient is therefore $64n^2C_*/\gamma-\gamma/2\le0$.  This is the entire self-play cancellation.  The best comparator $C$ in hindsight may lie on the boundary of $\cZ_i$,
where the log barrier diverges. To apply~\eqref{eq:rvu}, we approximate
$C$ by the strictly positive comparator
$C_\varepsilon=(1-\varepsilon)C+\varepsilon A_{c,i}$ with
$\varepsilon=1/(T+1)$.  Coordinatewise positivity gives barrier cost at most $m_i^2\log(T+1)$, while normalization bounds the cumulative smoothing loss by $2$.  Finally, the exact Gordon identity converts matrix regret back to linear $\Phi$-regret, and replacing the reset comparator by the original one costs at most $2\lambda T$.  These terms give~\eqref{eq:master}; see Appendix~\ref{app:closure} for the directed-divergence calculation.
\end{proof}

The theorem separates online optimization from game geometry: up to the comparator-radius factor, choosing the inverse step size $\gamma$ of order $n\sqrt{C_*}$ trades a barrier term of order $n\sqrt{C_*}(\sum_i m_i^2)\log(T+1)$ against the reset cost $2n\lambda T$.  Thus any class with a compact lift for all linear deviations and the fixed-point stability bound assumed in Theorem~\ref{thm:exact} inherits a fast self-play guarantee without a new regret decomposition.  Positivity constrains only the log-barrier iterates; smoothing recovers every boundary comparator.  We next supply two values of $C_*$: a reset-dependent one for arbitrary polytopes and a horizon-independent one for treeplexes.

\subsection{Arbitrary \texorpdfstring{$H$}{H}-polytopes: a weak reset}\label{sec:general-polytopes}
We suppress the player index and write $P$, $\cZ$, and $R$ for the corresponding positive slice, lift polytope, and entry-log barrier; $\|\cdot\|_P$ denotes the intrinsic strategy norm on $P$. Let $\Phi(P)$ denote the set of affine self-maps of $P$, and define
$\pi:\cZ\to\Phi(P)$ by $\pi(A)(s)=As$ for $s\in P$.

To compare reset fixed points, we associate each interior affine self-map
$z\in\Phi(P)$ with a law $\mathbb P_z$ on affine self-maps whose
coefficients average to those of $z$. Thus
$\mathbb E_{\phi\sim\mathbb P_z}[\phi(x)]=z(x)$ for every $x\in P$.
Starting at $q^0$, we stop with probability $\lambda$ before each step
and otherwise apply an independent draw from $\mathbb P_z$. The expected
endpoint is the reset fixed point of $z$, so we can compare two reset
fixed points through the laws of the sampled map sequences. By
independence, their KL divergence equals the divergence for a single
draw multiplied by the expected number of draws, $(1-\lambda)/\lambda$.
Our choice of laws makes the single-draw divergence a Bregman divergence
between the mean maps that the algorithm's entry-log barrier controls.

Increasing $\lambda$ thus improves the stability bound by shortening the
sequence, but it also gives more weight to $q^0$, incurring a cumulative
reset cost of order $\lambda T$. After tuning the learning rate, the
stability contribution scales as $\lambda^{-1/2}$ up to logarithmic and
game-size factors. Balancing the powers of $T$ gives
$\lambda\asymp T^{-2/3}$.

Represent an affine map by its matrix and translation in intrinsic coordinates. Let $\mu$ be normalized Lebesgue measure on this coefficient body $\Phi(P)$, of dimension $p_i\le p$ for player $i$, and define
\[
\Psi(\theta)=\log\mathbb E_{\phi\sim\mu}e^{\langle\theta,\phi\rangle},\qquad H(z)=\Psi^*(z):=\sup_\theta\{\langle\theta,z\rangle-\Psi(\theta)\}.
\]
For each interior $z\in\Phi(P)$, let $\mathbb P_z$ have density
\[
p_z(\phi)=e^{\langle\nabla H(z),\phi\rangle-\Psi(\nabla H(z))}
\]
relative to $\mu$. This exponential-family law has mean $z$, and its KL divergence agrees with the Bregman divergence between means:
\begin{equation}
\mathbb E_{\mathbb P_z}[\phi]=z,\qquad
D_H(z',z)=\KL(\mathbb P_{z'}\|\mathbb P_z).
\label{eq:entropy-lift}
\end{equation}
These laws are used only in the analysis: the algorithm neither samples from them nor evaluates $\Psi$ or $H$. The first lemma compares this auxiliary entropy geometry with the explicit log barrier; the second converts it into fixed-point stability.
\begin{lemma}[Entropy-to-log-barrier comparison]\label{lem:comparison}
For positive $A,B\in\cZ$, $D_H(\pi(B),\pi(A))\le4(p+1)^2D_R(B,A)$.
\end{lemma}
\begin{proof}[Proof sketch]
We scale a nonzero feasible tangent $V$ so that $\sum_{a,b}V_{ab}^2/A_{ab}^2=1$ and set $f(t)=H(\pi(A+tV))$.  Then $A+tV$ remains positive for $|t|<1$.  Self-concordance and the barrier-gradient inequality give $-p/(1+t)\le f'(t)\le p/(1-t)$ and $f''(t)\ge f''(0)/(1+\sqrt{f''(0)}|t|)^2$.  Integrating on $[-1/2,1/2]$ yields $f''(0)/(1+\sqrt{f''(0)}/2)\le4p$, hence $\sqrt{f''(0)}\le2p+2$ and $\nabla^2(H\circ\pi)\preceq4(p+1)^2\nabla^2R$.  A second integration along the segment from $A$ to $B$ gives the directed divergence bound; Appendix~\ref{app:comparison} records the endpoint and zero-direction details.
\end{proof}

\begin{lemma}[Reset stability]\label{lem:regenerative}
For $0<\lambda<1$ and positive $A,B\in\cZ$,
\begin{equation}
\|x_B^\lambda-x_A^\lambda\|_P^2
\le\frac{2(1-\lambda)(p+1)^2}{\lambda}D_R(B,A).
\label{eq:regen}
\end{equation}
\end{lemma}
\begin{proof}[Proof sketch]
Starting from $q^0$, compose a geometric number $K$ of independent affine maps drawn from $\mathbb P_{\pi(A)}$, with $\Pr(K=k)=\lambda(1-\lambda)^k$. Affinity and independence imply that the mean endpoint satisfies the reset equation, so it equals $x_A^\lambda$.

To compare $A$ and $B$, condition on $K=k$. The KL chain rule gives $kD_H(\pi(B),\pi(A))$ between the two records of sampled maps. Averaging over their common geometric length gives total KL divergence
\[
\mathbb E[K]D_H(\pi(B),\pi(A))
=\frac{1-\lambda}{\lambda}D_H(\pi(B),\pi(A)).
\]
The record $(K,\phi_1,\ldots,\phi_K)$ determines the endpoint by
applying the sampled maps in order to $q^0$. By the data-processing
inequality, the KL divergence between the endpoint laws under $B$ and $A$
is therefore at most $\frac{1-\lambda}{\lambda}D_H(\pi(B),\pi(A))$. For any affine $f:P\to[0,1]$, the difference of its expectations under the two endpoint laws is bounded in absolute value by their total variation distance. Taking the supremum over $f$ as in Section~\ref{sec:model} and applying Pinsker gives
\[
\|x_B^\lambda-x_A^\lambda\|_P^2
\le\frac{1-\lambda}{2\lambda}D_H(\pi(B),\pi(A)).
\]
Lemma~\ref{lem:comparison} then yields~\eqref{eq:regen}.
\end{proof}
The important feature is the single factor $1/\lambda$.  Direct subtraction of the reset equations gives $[I-(1-\lambda)B](x_B^\lambda-x_A^\lambda)=(1-\lambda)(B-A)x_A^\lambda$.  Nonexpansiveness bounds the inverse by $1/\lambda$, so a direct squared-sensitivity argument pays $1/\lambda^2$.  The regenerative/KL proof instead pays only the expected trajectory length $(1-\lambda)/\lambda$.  In the master bound this gives $C_*=\mathcal{O}(p^2/\lambda)$ and $\gamma\asymp n\sqrt{C_*}$, so we balance the leading terms $\gamma\log T+\lambda T$ at $\lambda\asymp T^{-2/3}$ and obtain $T^{1/3}$ regret.

\paragraph{Parameter choice and regret.}
For a known horizon $T$, set $a_T=\lceil(T+1)^{1/3}\rceil$ and choose
\begin{equation}
\lambda=a_T^{-2},\qquad \gamma=32n(p+1)a_T.
\label{eq:schedules-general}
\end{equation}
Then $C_*\le2(p+1)^2a_T^2$, so Theorem~\ref{thm:exact} yields the first line of Theorem~\ref{thm:informal}.  The chosen parameters satisfy $\gamma^2\ge128n^2C_*$, and substitution gives
\[
\sum_i\LinReg_i(T)\le32n(p+1)a_T\Bigl(\sum_i m_i^2\Bigr)\log(T+1)+\frac{1}{8(p+1)a_T}+2n(T/a_T^2+1),
\]
which is $\widetilde{\mathcal{O}}(\poly(D)T^{1/3})$.  Dividing by $T$ gives $\widetilde{\mathcal{O}}(\poly(D)T^{-2/3})$ maximum expected loss reduction from a linear deviation under the empirical distribution.

\subsection{Perfect-recall EFGs: stability without reset}\label{sec:efg-stability}
The logarithmic bound uses a property of the strategy treeplex beyond the polynomial size of its deviation representation. In terminal-sequence coordinates the treeplex carries a recursive norm in which all feasible matrices are nonexpansive. We use the recursion to prove the trace inequality below.  For its positive-slice representation $P$, let $\mathcal N(P)$ consist of the nonnegative \emph{mass functionals}, the weights that assign total mass one to every feasible plan. We use these weights to define the norm
\[
\mathcal N(P)=\{\ell\in\R^m_+:\ell^\top x=1\ \forall x\in P\},\qquad
\|z\|_{X_P}=\max_{\ell\in\mathcal N(P)}\ell^\top|z|,
\]
where $|z|$ is taken coordinatewise. This is the treeplex $\ell_1$ norm on terminal sequences of \citet[Lem.~4.2]{fan2024treeplex}, as the recursion below shows. It acts on the full coordinate space. The strategy norm $\|\cdot\|_P$ acts on the span of feasible differences. On the probability simplex, $\mathcal N(P)=\{\mathbf{1}\}$, so the two norms on feasible differences are $\|\cdot\|_1$ and $\tfrac12\|\cdot\|_1$, respectively. Nonnegativity and $A^\top\mathcal N(P)\subseteq\mathcal N(P)$ imply $\|Az\|_{X_P}\le\|z\|_{X_P}$ for every feasible deviation matrix $A$. Products and joins obey
\begin{equation}
\|z\|_{X_{P_1\times\cdots\times P_b}}=\max_j\|z_j\|_{X_{P_j}},\qquad
\|z\|_{X_{\operatorname{join}(P_1,\ldots,P_b)}}=\sum_j\|z_j\|_{X_{P_j}}.
\label{eq:norm-recursion}
\end{equation}
This exact sum/max recursion lets a sign propagate through the entire tree rather than multiplying a depth-dependent sensitivity constant.

By the Perron--Frobenius theorem, every entrywise positive square matrix
$A$ has a simple positive eigenvalue that exceeds the absolute value of
every other eigenvalue. This eigenvalue is called the \emph{Perron root}
and equals the spectral radius of $A$. Its positive right and left eigenvectors are the \emph{Perron vectors}, each unique up to scaling \citep[Sec.~8.2]{hornjohnson2013}. For a feasible $A>0$, choose a right Perron vector $v>0$. Its eigenvalue is positive, so it equals $\|Av\|_{X_P}/\|v\|_{X_P}$. Nonexpansiveness bounds this ratio by one. For the reverse bound, Proposition~\ref{prop:slack}(iii) gives a fixed point $s\in P$. Since $w^\top s=1$, the identity $As=s$ exhibits a nonzero eigenvector with eigenvalue $1$. The spectral radius of $A$ is therefore exactly $1$. We write $s_A$ and $l_A$ for the right and left Perron vectors, with
\[
A s_A=s_A,\quad w^\top s_A=1,\qquad
l_A^\top A=l_A^\top,\quad l_A^\top s_A=1.
\]
The learner plays $s_A$, so stability asks how this normalized right eigenvector changes with $A$.

A direct bound on the resolvent could depend on a spectral gap or a minimum transition probability, which may be arbitrarily small. The fixed-point argument needs only the sign of a single trace. The product and join recursion preserves that sign.

\begin{lemma}[Resolvent trace inequality]\label{lem:schur}
For any norm recursively built from absolute value using the two operations in~\eqref{eq:norm-recursion}, we write $\|T\|_{\mathrm{op}}:=\sup_{\|z\|\le1}\|Tz\|$ for its induced operator norm.  If $\|U\|_{\mathrm{op}}<1$ and $\|V\|_{\mathrm{op}}\le1$, then $\tr((I-U)^{-1}(I-V))\ge0$.
\end{lemma}
\begin{proof}[Proof sketch]
At a scalar leaf the trace is $(1-v)/(1-u)\ge0$. For a product or join, eliminate the complementary blocks by a Schur complement, leaving one child at a time. The sum or maximum norm ensures that the reduced $U$ is still a strict contraction and the corresponding reduced $V$ is nonexpansive. The block inverse formula then expresses the original trace as a sum of traces of the same form on the children, using cyclicity in the maximum case. Each is nonnegative by induction. Appendix~\ref{app:schur} gives the block formulas and norm bounds.
\end{proof}

For a positive feasible $A$ and a nonzero direction $V$ satisfying the homogeneous lift equalities, we set $\eta=\max_{a,b}|V_{ab}|/A_{ab}$, $A^\pm=A\pm V/\eta$, and $L(\alpha,\beta)=\alpha(I-A^+)+\beta(I-A^-)$.  The matrices $A^\pm$ are nonnegative by the choice of $\eta$ and have positive sum $2A$, so their weighted average $\widetilde A=(\alpha A^++\beta A^-)/(\alpha+\beta)$ is positive for $\alpha,\beta>0$. Because $L=(\alpha+\beta)(I-\widetilde A)$, the columns and rows of $\adj L$ lie in the right and left Perron eigenspaces of $\widetilde A$, respectively. Thus $\adj L$ is a multiple of $s_{\widetilde A}l_{\widetilde A}^\top$, which has trace one. This makes $\tr(\adj L)$ the normalization factor in the fixed-point formula (Appendix~\ref{app:trace-stability}).

For trigger deviations, \citet{anagnostides2023} use rational fixed-point formulas with positive coefficients. For linear deviations it suffices to show that the denominator is nondecreasing in each coefficient of $L$. The next lemma extracts precisely this property from Lemma~\ref{lem:schur}.
\begin{lemma}[Monotone Perron denominator]\label{lem:perron-denominator}
For $\alpha,\beta>0$, $a(\alpha,\beta):=\tr(\adj L(\alpha,\beta))$ is positive and satisfies $\partial_\alpha a,\partial_\beta a\ge0$.
\end{lemma}
\begin{proof}[Proof sketch]
For $\varepsilon>0$, put $c=\alpha+\beta+\varepsilon$ and $U=(\alpha A^++\beta A^-)/c$. Since $A^\pm\in\cZ$, both are nonexpansive in the treeplex norm
$\|\cdot\|_{X_P}$, as shown before~\eqref{eq:norm-recursion}.
Hence $\|U\|_{\mathrm{op}}\le(\alpha+\beta)/c<1$. Thus $I-tU$ is invertible for $0\le t\le1$, since $x=tUx$ forces $x=0$. Its determinant varies continuously from $1$ at $t=0$ and never vanishes, so $L+\varepsilon I=c(I-U)$ has positive determinant.

For a differentiable matrix $B(t)$ with positive determinant, Jacobi's formula gives
\[
\frac{d}{dt}\log\det B(t)=\tr\!\bigl(B(t)^{-1}B'(t)\bigr).
\]
Here we vary $\alpha$ with $\beta$ and $\varepsilon$ fixed. Then $\partial_\alpha(L+\varepsilon I)=I-A^+$ and $(L+\varepsilon I)^{-1}=c^{-1}(I-U)^{-1}$. Substituting these expressions and applying Lemma~\ref{lem:schur} with $V=A^+$ gives
\[
\partial_\alpha\log\det(L+\varepsilon I)
=c^{-1}\tr\bigl((I-U)^{-1}(I-A^+)\bigr)\ge0.
\]
The same argument applies to $\beta$. Thus $\det(L+\varepsilon I)/\varepsilon$ is nondecreasing in each coefficient, and its limit as $\varepsilon\to 0^+$ is $a(\alpha,\beta)$. The positive matrix $\widetilde A$ has simple Perron root $1$, so $L$ has a simple zero eigenvalue and this limit is strictly positive. Appendix~\ref{app:schur} gives the determinant argument.
\end{proof}

\begin{theorem}[Treeplex fixed-point stability]\label{thm:efg-stability}
For every treeplex with $m$ coordinates indexed by terminal sequences and positive $A,B\in\cZ$,
\begin{equation}
\|s_B-s_A\|_P^2\le200m^4D_R(B,A).
\label{eq:efg-stability}
\end{equation}
\end{theorem}
\begin{proof}[Proof sketch]
For $m\ge2$, set $r=m-1$ and let $s(t)=s_{A+tV/\eta}$ denote the
fixed point of the perturbed matrix. We bound $s'(0)$ to control the
sensitivity of the played strategy in this direction. Write $L_t=L(1+t,1-t)$. As shown in Appendix~\ref{app:trace-stability}, for $|t|<1$ the
normalized fixed point satisfies
\[
s(t)=\frac{\adj(L_t)s^0}{\rho(t)},\qquad
\rho(t):=\tr(\adj(L_t))=a(1+t,1-t)>0.
\]
Each entry of $\adj(L_t)$ is a signed $r\times r$ minor of $L_t$.
Since $L_t$ is affine in $t$, these entries and their trace $\rho(t)$
are polynomials of degree at most $r$. For any centered affine test
$f:P\to[-1/2,1/2]$, write $f(s)=g^\top s+b$. Then
\[
f(s(t))=\frac{F(t)}{\rho(t)},\qquad
F(t):=g^\top\adj(L_t)s^0+b\rho(t),\qquad \deg F\le r.
\]
By the monotonicity from Lemma~\ref{lem:perron-denominator} and the
degree-$r$ homogeneity of $a$, we have $|\rho'(0)|\le r\rho(0)$ and
$\rho(t)\le 2\rho(0)$ for $|t|\le 1/(2r)$. Since $|f|\le1/2$, this also bounds $|F(t)|$ by $\rho(0)$ on that interval. For a polynomial $p$ of degree at most $r$, Bernstein's inequality at the origin reads $|p'(0)|\le r\max_{|u|\le1}|p(u)|$ \citep{queffelec2019}. Taking $p(u)=F(u/(2r))$ gives $|F'(0)|\le2r^2\rho(0)$. The common factor $\rho(0)$ cancels in the quotient rule. Taking the supremum over centered affine tests gives
$\|s'(0)\|_P\le\tfrac52(m-1)^2$ for the path $s(t)=s_{A+tV/\eta}$.
Since $s_{A+tV}=s(\eta t)$, its derivative in the original direction
$V$ is $\eta s'(0)$. Taking norms and substituting the definition of $\eta$ gives 
\begin{equation}
\left\|\left.\frac{d}{dt}s_{A+tV}\right|_{t=0}\right\|_P
\le \tfrac52(m-1)^2\max_{a,b}\frac{|V_{ab}|}{A_{ab}}.
\label{eq:efg-local-main}
\end{equation}
For $D_R(B,A)<1/16$, the entrywise log barrier controls these relative changes along the segment from $A$ to $B$. Integrating~\eqref{eq:efg-local-main} gives~\eqref{eq:efg-stability}. For larger divergences the strategy diameter bound suffices. Appendix~\ref{app:trace-stability} gives the full proof.
\end{proof}
\paragraph{Parameter choice and regret.}
The stability bound gives $C_*=200M^4$ without a reset. Choose
\begin{equation}
\lambda=0,\qquad \gamma=320nM^2.
\label{eq:schedules-efg}
\end{equation}
These parameters satisfy Theorem~\ref{thm:exact} and give $\LinReg_i(T)\le320n^2M^4\log(T+1)+1/(80M^2)+2n$.  Hence the empirical distribution of play approaches linear correlated equilibrium at rate $\mathcal{O}(\poly(D)\log(T+1)/T)$.

The contrast with the generic argument is structural rather than merely quantitative.  In the general case stability is manufactured by a reset and its cost appears explicitly through $1/\lambda$; for treeplexes the exact unreset map is stable with $C_*=\mathcal{O}(M^4)$. The tree recursion preserves the trace inequality at every level, yielding
a stability bound that holds uniformly over all positive feasible deviation matrices. With $\lambda=0$, this
uniform bound lets us choose the inverse step size $\gamma$ independently
of $T$. The only term in~\eqref{eq:master} that grows with $T$ is then
the comparator term, which is proportional to $\log(T+1)$. Theorem~\ref{thm:exact} also applies to other compact representations for which one can establish a comparable stability estimate.

\paragraph{UTC implementation.}
The terminal-sequence matrix lift $\cZ_i$ describes a deviation by its action on realization plans. The UTC polytope $Z_i$ of \citet{zhang2024} describes the same linear deviations by the player's choices to act or to query the mediator. Its local flow constraints can be read directly from the game tree. ORBITS can optimize over $Z_i$ with the entrywise log barrier $\widehat R_i(z):=-\sum_e\log z_e$ and play the fixed point of the induced terminal-sequence map $C_i(z)$. The two barriers can select different iterates, so the UTC update needs its own stability bound.
\begin{corollary}[UTC implementation of ORBITS]\label{cor:utc-implementation}
Let $q_i$ be the number of coordinates in player $i$'s UTC log barrier after removing the fixed root row. For positive $z,z'\in Z_i$,
\[
\|s_{C_i(z')}-s_{C_i(z)}\|_i^2\le200M^4D_{\widehat R_i}(z',z).
\]
The UTC update competes with every linear deviation. If $\gamma^2\ge128n^2(200M^4)$, its exact iterates satisfy
\[
\sum_i\LinReg_i(T)\le\gamma\Bigl(\sum_iq_i\Bigr)\log(T+1)+4n/\gamma+2n.
\]
Eliminating one-action information sets gives $q_i\le5m_i^2$ and recovers the $\mathcal O(n^2M^4\log(T+1))$ rate with~\eqref{eq:schedules-efg}.
\end{corollary}
\begin{proof}[Proof sketch]
Each entry of $C_i(z)$ is a nonnegative linear combination of the UTC coordinates. Hence a feasible direction $h$ satisfies
\[
\max_{a,b}\frac{|C_i(h)_{ab}|}{C_i(z)_{ab}}
\le\max_e\frac{|h_e|}{z_e}
\le\sqrt{\nabla^2\widehat R_i(z)[h,h]}.
\]
Thus the UTC barrier controls the relative matrix perturbation in~\eqref{eq:efg-local-main}. Integrating that derivative bound gives stability in UTC coordinates. Every linear deviation has a UTC representative, so the same regret argument applies with $q_i$ barrier terms. Appendix~\ref{app:utc-implementation} gives the details.
\end{proof}
Either representation gives the same hindsight comparator class. Using UTC also lets ORBITS and UTC-CFR/RM+ learn on the same play and query variables in the experiments, where a common oracle computes the best linear deviation in hindsight (Appendix~\ref{app:experiments}). Section~\ref{sec:implementation} compares the per-round costs of the two ORBITS implementations.

\subsection{Per-iteration complexity}\label{sec:implementation}
The deviation update (line~3) solves one log-barrier-regularized convex program. For general polytopes we use the matrix lift $\cZ_i$. In EFGs the terminal-sequence matrix lift has $m_i^2$ barrier entries. The UTC implementation uses $q_i=(|\Sigma_i|-1)|\Sigma_i|+|\mathcal J_i|^2$ play and query coordinates (Appendix~\ref{app:utc-implementation}). Their local flow equations give a sparse constraint matrix for the Newton solver used in our experiments (Appendix~\ref{app:experiments}). Both implementations satisfy the polynomial per-round bit bound below.

The strategy step (line~4) solves $[I-(1-\lambda)A]x=\lambda q^0$ with reset, or $(I-A)x=0$ with $w_i^\top x=1$ without it. Full-information feedback is evaluated at the current compact profile and pulled back to deviation space. The cost update (line~6) and its addition to the cumulative score are linear in the lift coordinates.

For either domain let $k_i$ be the number of barrier entries ($m_i^2$ or $q_i$) and let $u^0>0$ be a rational feasible anchor. The prediction score has range at most $2(T+1)$, so first-order optimality bounds entry $e$ of the exact minimizer below by $u^0_e/[k_i+2(T+1)/\gamma]$. This floor has polynomial encoding length. For the exact score $b$ and minimizer $u^*$, the objective gap of a positive feasible $\widehat u$ is exactly $\gamma D_R(\widehat u,u^*)$. Appendix~\ref{app:implementation} uses this identity to control the effects of feedback rounding and approximate optimization on the profiles actually played. With the stated schedules their combined cost is less than one in the sum of regrets. Each round uses $\poly(D,\log(T+1))$ bit operations. Appendix~\ref{app:anytime} covers unknown horizons.

%% file: sections/conclusion.tex
\section{Conclusion}
ORBITS gives efficient uncoupled dynamics for minimizing linear $\Phi$-regret in full-information self-play. On a compact $H$-polytope, ORBITS mixes the learned deviation with the constant map to an interior strategy. This reset limits how far the played fixed point can move as the deviation changes. Balancing its stabilizing effect against the loss incurred by the reset gives $\widetilde{\mathcal{O}}(\poly(D)T^{1/3})$ cumulative regret against every affine self-map. Each round takes time polynomial in the compact description length and $\log(T+1)$.

In perfect-recall EFGs, the recursive structure of the sequence-form polytope gives direct control of the exact fixed point. The resulting $\mathcal{O}(n^2M^4\log(T+1))$ regret bound extends the logarithmic trigger-regret guarantee to all linear/UTC deviations. For a fixed game, the largest expected gain from any such deviation under the empirical distribution of play is therefore $\mathcal{O}(\log(T+1)/T)$. Whether efficient uncoupled dynamics can achieve polylogarithmic linear $\Phi$-regret on every compact $H$-polytope with polynomial dependence on its description length remains open.
\label{main-text-end}

%% file: appendices/related-work.tex
\section{Extended Related Work}\label{app:related}
\paragraph{Near-constant and constant swap regret.}
The adversarial full-swap row of Table~\ref{tab:comparison} reports a bound uniform in $N$ and $T$. The round requirement of \citet[Cor.~1.2]{dagan2024swap} has the form $(c\log(N)/\varepsilon^2)^{c/\varepsilon}$ for a universal $c$; \citet{pengrubinstein2024swap} obtain the same tradeoff. To invert it, put $L=\log T$ and $J=\log\log(eNT)$. For $\varepsilon=CJ/L<1$ we have $\log(c\log(N)/\varepsilon^2)=\mathcal O(J)$, since $\log\log N\le J$ and $\log L\le J$. Choosing a sufficiently large universal $C$ makes the round requirement at most $T$. When $CJ/L\ge1$, the trivial linear bound suffices. For $N,T\ge2$ this gives the displayed $\mathcal O(T\log\log(NT)/\log T)$.

The most recent normal-form advances refine the interaction between the Blum--Mansour reduction and optimistic learning.  \citet{tsuchiya2026} combine entropy and log-barrier regularization to improve the logarithmic horizon factor to its square root.  \citet{mai2026constant} instead control prediction errors for deviation gains through two-scale higher-order optimism, obtaining a horizon-independent bound while accounting for changes in the transition matrix's stationary distribution.  That work also gives a robust variant preserving the constant self-play bound up to a universal factor and guaranteeing $\mathcal O(\sqrt{NT\log N})$ swap regret against arbitrary bounded payoff sequences.  The adversarial guarantee still depends polynomially on the explicit action count, so it does not contradict the compact-domain lower bound discussed in Section~\ref{sec:related}.

The comparator-adaptive results of \citet{hait2025adaptive} give a distinct earlier route to constant regret under a restriction on the game: the sum of the players' \emph{unclipped external regrets} must be nonnegative for every play sequence.  Their bound specializes to $\mathcal{O}((nN+n^2)\log N)$ for full swap regret.  This condition does not hold in arbitrary general-sum games and should not be conflated with the nonnegativity of swap regret itself.  \citet{tsuchiya2026scale} address a different issue, obtaining logarithmic swap regret with fully scale-free, scale-invariant dynamics.

\paragraph{Constant external regret.}
Recent horizon-independent external-regret results include higher-order optimistic dynamics \citep{liu2026constant,abbadi2026constant}, multiplicative optimism \citep{soleymani2026constant}, and a horizon-independent analysis of Optimistic Hedge \citep{ha2026constant}.  The latter bound has an implicit dependence on the finite game's dimensions.  These guarantees concern the external comparator class; they do not by themselves yield constant swap regret or a compact algorithm for linear $\Phi$-regret.  Applying an external-regret method separately to the rows of a Blum--Mansour transition matrix requires an additional analysis, because the stationary distribution couples the row learners' feedback \citep{mai2026constant}.

\paragraph{Other compact approaches.}
Structured external-regret methods exploit sampling and kernelization to avoid pure-action enumeration in several game families \citep{beaglehole2023,farina2022kernel,kontogiannis2025kernel}.  Their equilibrium guarantees are based on external regret and need not control regret against every linear deviation.  \citet[Thm.~4]{cai2026proximal} bound individual proximal regret in Lipschitz, smooth convex games. To control regret against a symmetric affine deviation, they mix it with the identity to obtain a proximal map and rescale the resulting regret bound.  The entropy geometry used in the general-polytope proof is due to \citet{bubeckeldan2015}, with the exact self-concordance parameter established by \citet{chewi2021}.

%% file: appendices/polyhedral-geometry.tex
\section{Polyhedral Preliminaries and Intrinsic Geometry}\label{app:polyhedral}
\subsection{Affine-hull reduction and encoded geometry}
Suppress the player index and write $Q=\{x\in\R^d:Cx\le b\}$, where $C\in\R^{m\times d}$. We assume without loss of generality that $Q$ is full-dimensional and that no inequality is tight throughout $Q$.

Indeed, given any nonempty rational compact polytope, maximize each slack over it by linear programming (LP). The rows with maximum zero are equalities on the entire polytope. Their common equality space is its affine hull: average one feasible positive-slack witness for each remaining row to obtain a point at which all those rows are strict. A neighborhood of that point in the equality space is therefore feasible. If there are no remaining rows, the feasible set is the equality space itself. A rational parametrization $x=x^0+Uy$ of this space eliminates the equalities. In the new coordinates, loss differences use $U^\top\ell$ and conjugation by the parametrization gives a bijection of affine self-maps. Thus every deviation comparison, and hence regret, is preserved. These reductions use polynomial-time LP and rational linear algebra \citep{gls1988}. A zero-dimensional strategy set contributes zero regret.

Write $c_j^\top$ for row $j$ of $C$. Coordinatewise LP gives $Q\subseteq[-R,R]^d$ with rational $R\ge1$ of polynomial encoding length. An interior box is obtained from
\[
\max_{q,r}\ r\quad\text{subject to}\quad
c_j^\top q+r\|c_j\|_1\le b_j\quad(j\in[m]),\qquad 0\le r\le1.
\]
The constraints say precisely that $q+r[-1,1]^d\subseteq Q$. Full dimensionality gives a positive optimal $r$ and a rational optimal pair $(q,r)$ of polynomial encoding length.

Write an affine map as $x\mapsto Gx+h$. If each coefficient of $G$ and $h-q$ has magnitude at most $r/(dR+1)$, the image of every $x\in Q$ belongs to that interior box. Hence $\Phi(Q)$ contains a box in all $d(d+1)$ coefficient directions. For an outer bound, evaluate any self-map at $q$ and $q+re_j$, where $e_j$ is coordinate vector $j$. Both images lie in $[-R,R]^d$, so each entry of $Ge_j$ has magnitude at most $2R/r$. The identity $h=\phi(q)-Gq$ then bounds $\|h\|_\infty\le R+2dR^2/r$. These arguments give inner and outer radii with polynomial \emph{encoding length}, sufficient even if $r$ is exponentially small.

\subsection{Affine-map inequalities and a compact nonnegative lift}
Row-wise LP duality characterizes the affine maps in original coordinates:
\begin{equation}
(G,h)\in\Phi(Q)
\quad\Longleftrightarrow\quad
\exists\Lambda\in\R_+^{m\times m}:\ \Lambda C=CG,\quad \Lambda b+Ch\le b.
\label{eq:farkas-endo}
\end{equation}
Indeed, row $j$ requires $\max_{Cx\le b}c_j^\top Gx+c_j^\top h\le b_j$. Its dual minimizes $b^\top\upsilon_j+c_j^\top h$ over $\upsilon_j\ge0$ satisfying $C^\top\upsilon_j=G^\top c_j$. The primal is feasible and bounded, so dual optimal solutions exist. Stacking their transposes gives $\Lambda$. The multiplier fibers may be unbounded even though $\Phi(Q)$ is compact.

For the online barrier set $s=b-Cx$, $s^0=b-Cq>0$, and $W=\operatorname{range}(-C)$. If $Cv=0$ with $v\ne0$, a line through any feasible point would lie in $Q$, contradicting boundedness. Thus the slack map is an affine isomorphism onto $P=(s^0+W)\cap\R_+^m$, where $\dim W=d$ and $\aff P=s^0+W$. Also $W\cap\R_+^m=\{0\}$, since any nonzero vector in this intersection would give an unbounded ray from $s^0$. The strict theorem of alternatives therefore gives $w>0$ perpendicular to $W$. Scale it so $w^\top s^0=1$, whence $w^\top s=1$ on $\aff P$. A rational choice is obtained by maximizing $t$ subject to $C^\top w=0$, $w^\top s^0=1$, and $w_a\ge t\ge0$ for every $a$.

For an EFG, Appendix~\ref{app:efg-prelims} constructs a nonnegative linear map reconstructing the full plan from its terminal sequences. On the affine hull of the terminal projection the reconstructed coordinates satisfy the sequence-form equalities by affinity. Nonnegative terminal coordinates reconstruct a nonnegative plan and reproduce the given terminal entries. Thus the terminal polytope also equals its affine hull intersected with the nonnegative orthant. Uniform positive behavioral probabilities give a positive reference point.

\begin{proof}[Proof of Proposition~\ref{prop:slack}]
\emph{(i) Compactness.} Let the rows of $F$ span $W^\perp$, let the columns of $B$ span $W$, and put $h_0=Fs^0$. Then $P=\{s\ge0:Fs=h_0\}$. Preserving $\aff P$ is equivalent to $FAs^0=h_0$ and $FAB=0$. These equalities and $A\ge0$ define the closed lift $\cZ=\cZ(P)$ of~\eqref{eq:slack-lift}. Since $As^0\in P$, nonnegativity implies
\[
A_{ab}s_b^0\le(As^0)_a\le\max_{s\in P}s_a.
\]
Every coordinate is therefore bounded. The matrix $s^0w^\top$ is strictly positive and sends the entire affine hull to $s^0$.

\emph{(ii) Representation.} For each output coordinate of an affine self-map, choose an ambient affine representation $g^\top s+\beta$ on $P$. Its minimum is nonnegative. By strong LP duality there is an unrestricted multiplier $y$ with $F^\top y\le g$ and
\[
\kappa:=h_0^\top y+\beta
=\min_{Fs=h_0,\ s\ge0}(g^\top s+\beta)\ge0.
\]
On $P$ this output equals $(g-F^\top y+\kappa w)^\top s$, whose coefficient vector is nonnegative. Stacking the rows gives a nonnegative matrix agreeing with the map on $P$. Its linearity and $A(P)\subseteq P$ imply $A(\aff P)\subseteq\aff P$, as required. Conversely, every feasible matrix preserves both the affine slice and nonnegativity, hence maps $P$ into $P$.

\emph{(iii) Fixed points.} Brouwer's theorem gives each feasible $A$ a fixed point in the nonempty compact convex set $P$. Every point of $P$ satisfies $w^\top s=1$ and is nonzero. Moreover, $A^k$ belongs to the bounded lift for every integer $k\ge1$, so no eigenvalue has modulus greater than one. For $A>0$, Perron--Frobenius gives a simple eigenvalue one, strictly positive right and left eigenvectors, and all other eigenvalues of modulus strictly below one. The feasible fixed point is the unique right Perron vector $s_A$ normalized by $w^\top s_A=1$. Normalize the left eigenvector $l_A$ by $l_A^\top s_A=1$. For any $s\in P$, the set $\{x\in P:l_A^\top x=l_A^\top s\}$
is nonempty, compact, and convex. Since $l_A^\top A=l_A^\top$,
$A$ maps this set into itself. By Brouwer's theorem it contains
a fixed point, which must be $s_A$ by uniqueness. Hence
$l_A^\top s=l_A^\top s_A=1$ for every $s\in P$. This last identity will be used in Appendix~\ref{app:trace-stability}.
\end{proof}

Write $\pi:\cZ\to\Phi(P)$ for restriction to $\aff P$ in intrinsic affine-map coordinates. This is an affine surjection by part (ii). Its linear part maps the direction space of the lift onto the direction space of $\Phi(P)$, since it maps affine hulls onto affine hulls. A sufficiently small relative neighborhood of any positive $A$ lies in $\cZ$. A surjective linear map in finite dimensions maps relative neighborhoods to neighborhoods of their images, so $\pi(A)$ is an interior point of the coefficient body. This is the interior property used in the entropy comparison.

\subsection{Nonexpansiveness and payoff normalization}
The linear part $g$ of an affine test $f:P\to[0,1]$ satisfies $|g(u-v)|\le1$ for all $u,v\in P$. Conversely, every linear functional on $W$ satisfying these inequalities defines such a test: subtract the minimum of $s\mapsto g(s-s^0)$ on $P$. Thus the set of test functionals is the polar of $P-P$ in $W$. The difference set is compact, convex, symmetric and contains a neighborhood of zero in $W$. By the bipolar theorem it is the unit ball of the strategy norm, whose formula on $W$ is
\[
\|v\|_P=\inf\{a>0:v=a(x-y),\ x,y\in P\}.
\]
For $v\ne0$ the infimum is attained by compactness. If $v=a(x-y)$, the linear part of any affine self-map sends $v$ to $a(\phi(x)-\phi(y))$, so it has operator norm at most one. Thus $(1-\lambda)\phi+\lambda q^0$ has Lipschitz constant at most $1-\lambda$ for $0<\lambda<1$ and has a unique fixed point in $P$. Likewise, evaluating a difference of two feasible deviations at a point of $P$ gives an element of $P-P$.

Since $\|g\|_{P,*}=\max_{u,v\in P}|\langle g,u-v\rangle|$, losses in $[-1,1]$ yield $\|\ell_i(x_{-i})\|_{i,*}\le2$. To compare two opponents' profiles, change one opponent $j$ at a time. If $a=\|x_j-y_j\|_j>0$, write $x_j-y_j=a(u_j-v_j)$ for some $u_j,v_j\in P_j$. For two own strategies $u_i,v_i$, the induced change in their loss difference is $a$ times
\[
L_i(u_i,u_j,\ldots)-L_i(v_i,u_j,\ldots)
-L_i(u_i,v_j,\ldots)+L_i(v_i,v_j,\ldots)
\]
by multi-affinity, with other players fixed. The absolute value of the displayed expression is at most four. A zero difference contributes zero. Taking the supremum over $u_i,v_i$, summing over opponents, and using Cauchy--Schwarz gives
\[
\|\ell_i(x_{-i})-\ell_i(y_{-i})\|_{i,*}
\le4\sum_{j\ne i}\|x_j-y_j\|_j
\le4\sqrt{(n-1)\sum_{j\ne i}\|x_j-y_j\|_j^2}.
\]
This proves~\eqref{eq:smoothness}, including the one-player case where the sum is empty.

%% file: appendices/treeplex-representation.tex
\section{Extensive-Form and Treeplex Representation}\label{app:efg-prelims}
We use the usual explicit, finite perfect-recall EFG model. At each information set, every node has the same action set. A player's own sequence records its earlier information sets and actions. Perfect recall makes the parent own sequence $p(j)$ well-defined at every information set $j$. The realization constraints are~\eqref{eq:sequence}; a nonzero-parent realization yields the behavioral probability $x_{ja}/x_{p(j)}$, and behavior at a zero-parent information set can be chosen arbitrarily. Conversely, multiplying own behavioral probabilities along each own sequence gives a feasible realization plan. Perfect recall ensures realization equivalence with mixed strategies \citep{koller1996}. Chance and the other players' reach probabilities occur in payoffs rather than in the player's realization constraints.

At a choice of the player with incoming mass one, $t_a$ is the realization weight assigned to action $a$. If the continuation strategy after $a$ lies in $P_a$, its coordinates are scaled by $t_a$. In disjoint coordinate blocks this gives the convex join
\[
\operatorname{join}(P_1,\ldots,P_b)
=\{(t_1z_1,\ldots,t_bz_b):t\in\Delta_b,\ z_j\in P_j\ (j\in[b])\},
\]
where $\Delta_b$ is the probability simplex. After an own sequence, different next information sets may arise through chance or opponents' actions. A plan specifies a continuation at each such information set, so the corresponding blocks form a Cartesian product. Starting from scalar leaves $\{1\}$ gives the recursive treeplex representation \citep{hoda2010treeplex,fan2024treeplex}.

After projection the remaining coordinates are indexed by terminal sequences. For example, if two information sets can follow the same own sequence of weight $t$, each of their outgoing action blocks sums to $t$. Either sum recovers $t$ and so does their average. In general, to recover a discarded incoming weight, take a child with mass functional $w_j^\top s=1$ and apply it to the scaled block, giving $w_j^\top(t_jz_j)=t_j$. At a product, average the positive child mass functionals. At a join, sum them. Recursing recovers the full realization plan by a linear map. Thus restriction to terminal sequences preserves all linear deviations and their regret.

The comparator class in this paper is the set of all affine self-maps $\phi_i:Q_i\to Q_i$ on realization plans. Adding the root coordinate $x_\varnothing=1$ makes every such map linear; conversely, a linear endomorphism on that affine slice is an affine self-map. Hence this representation captures the full linear/UTC deviation class in sequence form, while remaining distinct from arbitrary functions defined directly on pure contingent plans.

\subsection{Normalization duality and the common ambient norm}
For completeness, we restate the only auxiliary norm used in the EFG argument.  For the positive-slice representation $P\subset\R_+^m$ indexed by terminal sequences, let
\[
\mathcal N(P)=\{\ell\ge0:\ell^\top s=1\text{ for all }s\in P\},
\qquad
\norm{z}_{X_P}=\max_{\ell\in\mathcal N(P)}\ell^\top|z|,
\]
with $|z|$ taken coordinatewise. Since $\aff(P)=s^0+W$, a nonnegative weight $\ell$ belongs to $\mathcal N(P)$ exactly when $\ell\in W^\perp$ and $\ell^\top s^0=1$.
Positivity of $s^0$ gives $\ell_j\le1/s_j^0$ for every $\ell\in\mathcal N(P)$, and the mass functional above supplies a positive member. Moreover, $A\in\cZ(P)$ implies $A^\top\ell\in\mathcal N(P)$ for every $\ell\in\mathcal N(P)$. Hence
\[
\norm{Az}_{X_P}
\le\max_{\ell\in\mathcal N(P)}(A^\top\ell)^\top|z|
\le\norm z_{X_P}.
\]
This uses the same full lift; it does not extend the strategy polytope or alter its deviations.

For a Cartesian product $P=\prod_{j=1}^bP_j$, we write $\Delta_b:=\{t\in\R_+^b:\sum_{j=1}^b t_j=1\}$.  Varying one component at a time shows that $\ell_j^\top s_j=t_j$ is constant on each block. These constants are nonnegative and sum to one. If $t_j>0$ divide by $t_j$ to get a child mass functional. If $t_j=0$, testing the strictly positive child anchor forces the entire nonnegative block $\ell_j$ to vanish. Conversely, any choice of normalized blocks with these weights gives total mass one. Hence
\[
\mathcal N(P)
=\left\{(t_1\ell_1,\ldots,t_b\ell_b):
 t\in\Delta_b,\ \ell_j\in\mathcal N(P_j)\ \text{for }j=1,\ldots,b\right\}.
\]
For the convex join $P=\operatorname{join}(P_1,\ldots,P_b)$, testing points supported on a single child gives
\[
\mathcal N(P)=\prod_{j=1}^b\mathcal N(P_j).
\]
Maximizing $\ell^\top|z|$ in these two explicit formulas yields~\eqref{eq:norm-recursion}. At a scalar point the norm is absolute value. The recursion can have arbitrary depth; the Schur argument below propagates a sign rather than a dimension-dependent factor.

%% file: appendices/barrier-comparison.tex
\section{Proof of the Entropy-to-Log-Barrier Comparison}\label{app:comparison}

The following lemma is stated in a slightly more general form to isolate the new comparison from the particular entropy used later.

Recall that a $\nu$-self-concordant barrier on a full-dimensional convex body $K$ is a convex $C^3$ function $F:\operatorname{int}K\to\R$ with positive-definite Hessian and $F(x)\to+\infty$ as $x$ approaches $\partial K$ from the interior, such that every line restriction $f(t):=F(x+tv)$ satisfies
\[
|f'''(t)|\le2f''(t)^{3/2},\qquad
f'(t)^2\le\nu f''(t)
\]
whenever $x+tv\in\operatorname{int}K$. These are the self-concordance and barrier-gradient inequalities, respectively. The entropic barrier of a convex body with nonempty interior in $\R^k$ has $\nu=k$ \citep{bubeckeldan2015,chewi2021}. Additive normalizing constants in $F$ do not affect either inequality.

\begin{lemma}[Comparison on a positive lift]\label{lem:sc-comparison}
We consider a compact set $Z=\{z\ge0:Ez=b\}$ with a positive point, an affine surjection $\pi:Z\to K$ onto a full-dimensional compact convex body, and a $\nu$-self-concordant barrier $F$ on $K$ with $\nu\ge1$.  We write $R(z):=-\sum_a\log z_a$.  For positive $z,z'\in Z$,
\[
D_F(\pi(z'),\pi(z))\le4(\nu+1)^2D_R(z',z).
\]
\end{lemma}
\begin{proof}
We fix a positive $z$ and a direction $v$ with $Ev=0$. If $v=0$, the claim is immediate. We rescale $v$ so that $\sum_a v_a^2/z_a^2=1$. Since this implies $|v_a|\le z_a$ for every coordinate, the point $z+tv$ is positive and feasible for every $|t|<1$. It lies in $\ri Z$, so its image lies in $\operatorname{int}K$ by affine surjectivity (the neighborhood argument in Appendix~\ref{app:polyhedral}).  We write $f(t):=F(\pi(z+tv))$; if the projected direction is zero there is nothing to prove, so we assume $f''(0)>0$.

We first derive the endpoint gradient bounds
\begin{equation}
-\frac{\nu}{1+t}\le f'(t)\le\frac{\nu}{1-t}\qquad(-1<t<1).
\label{eq:line-gradient}
\end{equation}
If $f'(t)>0$, convexity makes it positive to the right. The barrier-gradient inequality gives $(1/f')'=-f''/(f')^2\le-1/\nu$. If $f'(t)>\nu/(1-t)$, integration would force $1/f'$ to be nonpositive strictly before $1$, contradicting finiteness and differentiability in the domain. This proves the upper bound, including the trivial nonpositive case. Reflecting the line proves the lower bound.

Self-concordance also gives $|(f''^{-1/2})'|\le1$.  We combine this differential estimate with~\eqref{eq:line-gradient} as
\begin{align*}
f''(t)
&\ge\frac{f''(0)}{\bigl(1+\sqrt{f''(0)}|t|\bigr)^2},\\
\frac{f''(0)}{1+\sqrt{f''(0)}/2}
&\le f'(1/2)-f'(-1/2)\le4\nu,\\
\sqrt{f''(0)}
&\le\nu+\sqrt{\nu^2+4\nu}\le2\nu+2.
\end{align*}
The first line follows by integrating $|(f''^{-1/2})'|\le1$.  The second integrates the resulting lower bound on $f''$ over $[-1/2,1/2]$ and applies~\eqref{eq:line-gradient}; the last line solves the resulting quadratic inequality.
Restoring the scale of $v$ proves pointwise Hessian domination,
\begin{equation}
\nabla^2(F\circ\pi)(z)[v,v]
\le4(\nu+1)^2\nabla^2R(z)[v,v].
\label{eq:hessian-domination}
\end{equation}
For $v=z'-z$, integrate~\eqref{eq:hessian-domination} along the positive feasible segment.  Since $\pi$ is affine,
\[
D_F(\pi(z'),\pi(z))
=\int_0^1(1-t)\,\nabla^2(F\circ\pi)(z+tv)[v,v]dt.
\]
This proves the result with the stated directed orientation.
\end{proof}

\begin{proof}[Proof of Lemma~\ref{lem:comparison}]
We apply Lemma~\ref{lem:sc-comparison} to the compact full-affine lift $\cZ$, its projection onto actual affine-map coefficients, and the entropic barrier on $\Phi(P)$, with $\nu=p_i=d_i(d_i+1)$ for player $i$. The coefficient body has nonempty interior by Appendix~\ref{app:polyhedral}. Since $p_i\le p$, the uniform comparison factor is $4(p+1)^2$.
\end{proof}

\paragraph{Geometric consequence.}
The entry-log barrier $R$ and the entropic barrier $H$ induce different geometries and generally have different minimizers. Lemma~\ref{lem:sc-comparison} controls the entropic divergence between projected barrier iterates by the explicit matrix divergence, with a polynomial loss in the affine-map dimension. This comparison supplies the generic reset-stability estimate used in Lemma~\ref{lem:regenerative}.

%% file: appendices/reset-stability.tex
\section{Regenerative Fixed-Point Stability}\label{app:regenerative}
\subsection{The entropy/KL identities}
We consider a full-dimensional compact convex body $\Phi\subset\R^k$ of affine-map coefficients, use its uniform base distribution $\mu$, and define
\[
\Psi(\theta)=\log\E_{\phi\sim\mu}\!\left[e^{\ip\theta\phi}\right],\qquad H=\Psi^*.
\]
Differentiating under the integral gives the mean of the tilted law as $\nabla\Psi(\theta)$ and its covariance as $\nabla^2\Psi(\theta)$. The latter is positive definite because the density is positive on a full-dimensional set. To see that every interior mean $z$ occurs, take $\rho>0$ with $z+2\rho B_2\subseteq\Phi$, where $B_2$ is the Euclidean unit ball. For $\theta\ne0$, integrate over the ball of radius $\rho/2$ about $z+\rho\theta/\|\theta\|_2$. Its $\mu$-mass $\alpha>0$ is independent of the direction of $\theta$, so
\[
\Psi(\theta)-\langle\theta,z\rangle\ \ge\ \tfrac{\rho}{2}\|\theta\|_2+\log\alpha.
\]
The strictly convex objective therefore has a unique minimizer $\theta_z$ with $\nabla\Psi(\theta_z)=z$. Convex conjugacy gives $\theta_z=\nabla H(z)$ and $H(z)=\langle\theta_z,z\rangle-\Psi(\theta_z)$ \citep{bubeckeldan2015,wainwrightjordan2008}. Thus, for each interior $z$, the pointwise density
\[
p_z(\phi):=\exp\!\left(\ip{\nabla H(z)}{\phi}-\Psi(\nabla H(z))\right),
\qquad \phi\in\Phi,
\]
is normalized relative to $\mu$ and has mean $z$.  We write $\mathbb P_z$ for the probability law obtained by weighting $\mu$ by $p_z$.  For all interior $z,z'$, direct substitution gives
\begin{align*}
\KL(\mathbb P_{z'}\|\mathbb P_z)
&=\ip{\nabla H(z')-\nabla H(z)}{z'}
-\Psi(\nabla H(z'))+\Psi(\nabla H(z))\\
&=D_H(z',z).
\end{align*}
This proves~\eqref{eq:entropy-lift}.  The laws $\mathbb P_z$ are used only in the analysis; the algorithm never samples from them.

\subsection{Proof of Lemma~\ref{lem:regenerative}}
\begin{proof}
We fix an interior mean affine map $z$ and write $x_z^\lambda$ for the unique solution of $x=(1-\lambda)z(x)+\lambda q^0$.  Starting from $q^0$, we draw a random number $K$ with
\[
\Pr(K=k)=\lambda(1-\lambda)^k,\qquad k=0,1,\ldots.
\]
Conditional on $K=k$, we sample independent affine maps $\phi_1,\ldots,\phi_k\sim\mathbb P_z$ and set
\[
X_z:=\phi_k\circ\cdots\circ\phi_1(q^0),
\]
with $X_z=q^0$ when $k=0$. Put $u_k=\E[X_z\mid K=k]$. Conditional on $K=k$, the last map is independent of the preceding $k-1$ maps; averaging first over that last map and then over its input gives $u_k=z(u_{k-1})$, with $u_0=q^0$. Hence the geometric mixture has mean
\[
\E X_z=\lambda q^0+(1-\lambda)\sum_{k\ge0}\lambda(1-\lambda)^k z(u_k)
      =\lambda q^0+(1-\lambda)z(\E X_z).
\]
The series converges because $P$ is compact, and the last equality uses affinity and that the weights sum to one. Uniqueness of the reset fixed point gives $\E X_z=x_z^\lambda$.

For two mean maps $z,z'$, we condition on the common event $K=k$.  Independence and the KL chain rule show that the conditional KL divergence from the $z'$-record law to the $z$-record law is
$k\,\KL(\mathbb P_{z'}\|\mathbb P_z)=k\,D_H(z',z)$, where the last equality is~\eqref{eq:entropy-lift}.  The record includes $K$, whose marginal law is the same in both experiments. The KL chain rule and $\E K=(1-\lambda)/\lambda$ thus give divergence $(1-\lambda)D_H(z',z)/\lambda$ between the complete trajectory records.  We write $Q_z$ for the distribution of the terminal point $X_z$.  Since $X_z$ is a function of the trajectory record, data processing gives
\[
\KL(Q_{z'}\|Q_z)\le\frac{1-\lambda}{\lambda}D_H(z',z).
\]
For every affine test $f:P\to[0,1]$, affinity gives $f(x_z^\lambda)=\E f(X_z)$, and boundedness gives
$|\E f(X_{z'})-\E f(X_z)|\le\TV(Q_{z'},Q_z)$.  Taking the supremum over such $f$ and applying Pinsker's inequality yields
\[
\norm{x_{z'}^\lambda-x_z^\lambda}_P^2
\le\frac{1-\lambda}{2\lambda}D_H(z',z).
\]
Finally, we set $z'=\pi(B)$ and $z=\pi(A)$ and apply Lemma~\ref{lem:comparison}.
\end{proof}

%% file: appendices/resolvent-trace.tex
\section{The Resolvent Trace Inequality}\label{app:schur}
\begin{proof}[Proof of Lemma~\ref{lem:schur}]
We argue by induction on the recursive sum/max construction of the norm. For the real line, $|u|<1$ and $|v|\le1$ imply $(1-v)/(1-u)\ge0$. Unary sums may be suppressed. For the inductive step, $X$ is an $\ell_1$ or $\ell_\infty$ sum of smaller spaces $X_1,\ldots,X_b$ on which the assertion holds. We write $r=\norm{U}_{\mathrm{op}}<1$. Fix a block $i$ and group the other blocks into $J$, with their inherited sum or maximum norm. Subscripts on operators specify their output and input blocks, in that order. Both coordinate injection and coordinate projection have operator norm one, hence $\norm{U_{JJ}}_{\mathrm{op}}\le r$.

For either $W=U$ or $W=U_{JJ}$, the bound $\norm{W^k}_{\mathrm{op}}\le r^k$ implies that the matrix geometric series $\sum_{k=0}^{\infty}W^k$ converges in operator norm. Multiplication of its partial sum by $I-W$ on either side gives
\[
 (I-W)\sum_{k=0}^{K}W^k
 = I-W^{K+1}
 = \left(\sum_{k=0}^{K}W^k\right)(I-W).
\]
Since $W^{K+1}\to0$ as $K\to\infty$, this proves the Neumann series formula $(I-W)^{-1}=\sum_{k=0}^{\infty}W^k$. We can therefore define
\[
B=(I-U)^{-1},\qquad D_i=(I-U_{JJ})^{-1},\qquad U^{[i]}=U_{ii}+U_{iJ}D_iU_{Ji}.
\]
Eliminating the $J$-block in $(I-U)(x,y)=(h,k)$ first gives $y=D_i(k+U_{Ji}x)$ and then $(I-U^{[i]})x=h+U_{iJ}D_i k$. Reading off the solution operators gives
\begin{equation}
B_{ii}=(I-U^{[i]})^{-1},\quad
B_{iJ}=B_{ii}U_{iJ}D_i,\quad B_{Ji}=D_iU_{Ji}B_{ii}.
\label{eq:schur-blocks}
\end{equation}
We first show $\norm{U^{[i]}}_{\mathrm{op}}\le r$. For $x\in X_i$ let $y=D_iU_{Ji}x$. Then $U(x,y)=(U^{[i]}x,y)$. In an $\ell_1$ sum,
\[
\norm{U^{[i]}x}+\norm y\le r(\norm x+\norm y),
\]
which gives the claim. In an $\ell_\infty$ sum, the contraction $y\mapsto U_{Ji}x+U_{JJ}y$ maps the ball of radius $\norm x$ to itself, since the full input $(x,y)$ then has norm at most $\norm x$. Its fixed point satisfies $\norm y\le\norm x$, and the $i$-th block row of $U$ gives $\norm{U^{[i]}x}\le r\norm x$.

For an $\ell_1$ sum, put
\[
V^{[i]}=V_{ii}+U_{iJ}D_iV_{Ji}.
\]
We replace the $i$-th block column of $U$ by the $i$-th block column of $V$, obtaining $Z$. Each block column is a contraction from its input block into $X$. Indeed, for $z=(z_k)_k$ the triangle inequality gives $\norm{Zz}\le\sum_k\norm{Z_{\cdot k}z_k}\le\sum_k\norm{z_k}=\norm z$. Taking $y=D_iV_{Ji}x$ yields $Z(x,y)=(V^{[i]}x,y)$, whence
\[
\norm{V^{[i]}x}+\norm y\le\norm x+\norm y.
\]
Thus $\norm{V^{[i]}}_{\mathrm{op}}\le1$. Formula~\eqref{eq:schur-blocks} gives
\[
[B(I-V)]_{ii}=(I-U^{[i]})^{-1}(I-V^{[i]}).
\]
Its trace is nonnegative by induction. Summing the diagonal-block traces proves the result in the $\ell_1$ case.

For an $\ell_\infty$ sum, instead define
\[
V^{[i]}=V_{ii}+V_{iJ}D_iU_{Ji}.
\]
The harmonic extension $y=D_iU_{Ji}x$ has norm at most $\norm x$, so applying block row $i$ of $V$ proves $\norm{V^{[i]}x}\le\norm x$. By cyclicity, $\tr(B(I-V))=\tr((I-V)B)$. The $i$-th diagonal block of the latter product is
\[
[(I-V)B]_{ii}=(I-V^{[i]})(I-U^{[i]})^{-1}.
\]
Its trace equals the inductive nonnegative expression on $X_i$. Summing proves the result. No factor depending on the number of sum/maximum levels appears.
\end{proof}

\subsection{Proof of Lemma~\ref{lem:perron-denominator}}
\begin{proof}
We fix $\alpha,\beta>0$ and define
\[
\widetilde A(\alpha,\beta)=\frac{\alpha A^++\beta A^-}{\alpha+\beta},\qquad
L(\alpha,\beta)=(\alpha+\beta)\bigl(I-\widetilde A(\alpha,\beta)\bigr).
\]
The matrices $A^\pm$ are feasible and nonnegative. Since $A^++A^-=2A>0$ entrywise and both coefficients are positive, their convex combination $\widetilde A(\alpha,\beta)$ is strictly positive. It is feasible, so Proposition~\ref{prop:slack} makes $1$ its simple Perron eigenvalue.  Hence $L(\alpha,\beta)$ has a simple zero eigenvalue and
\[
a(\alpha,\beta)=\tr(\adj L(\alpha,\beta))\ne0.
\]
We first establish the sign of its partial derivatives without differentiating at the singular matrix $L(\alpha,\beta)$.

For $\varepsilon>0$,
\[
L(\alpha,\beta)+\varepsilon I
=(\alpha+\beta+\varepsilon)
\left(I-\frac{\alpha+\beta}{\alpha+\beta+\varepsilon}\widetilde A(\alpha,\beta)\right).
\]
The scalar $(\alpha+\beta)/(\alpha+\beta+\varepsilon)$ lies in $(0,1)$.  Because $\widetilde A(\alpha,\beta)$ is positive, its Perron eigenvalue is $1$ and every other eigenvalue has modulus strictly below one.  Hence every real factor $1-\frac{\alpha+\beta}{\alpha+\beta+\varepsilon}\lambda_j$ is positive, while each nonreal conjugate pair contributes $\left|1-\frac{\alpha+\beta}{\alpha+\beta+\varepsilon}\lambda_j\right|^2>0$.  Therefore $\det\!\bigl(L(\alpha,\beta)+\varepsilon I\bigr)>0$.  Since every feasible deviation matrix is nonexpansive in the treeplex norm, Lemma~\ref{lem:schur} applied to
\[
U=\frac{\alpha+\beta}{\alpha+\beta+\varepsilon}\widetilde A(\alpha,\beta),
\qquad
V=A^+
\]
gives
\[
\tr\!\left[
\left(I-\frac{\alpha+\beta}{\alpha+\beta+\varepsilon}\widetilde A(\alpha,\beta)\right)^{-1}(I-A^+)
\right]\ge0.
\]
With $\beta$ and $\varepsilon$ fixed, $\partial_\alpha(L+\varepsilon I)=I-A^+$. Jacobi's formula gives $\partial_\alpha\log\det(L+\varepsilon I)=\tr((L+\varepsilon I)^{-1}(I-A^+))$. Substituting the inverse and using the trace bound above yields
\begin{equation}
\frac{\partial}{\partial \alpha}\log\det\!\bigl(L(\alpha,\beta)+\varepsilon I\bigr)
=\frac{1}{\alpha+\beta+\varepsilon}
\tr\!\left[
\left(I-\frac{\alpha+\beta}{\alpha+\beta+\varepsilon}\widetilde A(\alpha,\beta)\right)^{-1}(I-A^+)
\right]\ge0.
\label{eq:regularized-det-p}
\end{equation}
The same argument with $A^-$ gives
\begin{equation}
\frac{\partial}{\partial \beta}\log\det\!\bigl(L(\alpha,\beta)+\varepsilon I\bigr)\ge0.
\label{eq:regularized-det-q}
\end{equation}

For each fixed $\varepsilon>0$, the determinant is positive, so the two inequalities prove that $\varepsilon^{-1}\det(L+\varepsilon I)$ is nondecreasing in each coefficient. The determinant is a polynomial in $\varepsilon$. Its constant term vanishes, and the coefficient of $\varepsilon$ is the sum of the principal $(m-1)$-minors, which equals $\tr(\adj L)$. Therefore
\[
  \lim_{\varepsilon\to 0^+}\frac{\det(L+\varepsilon I)}{\varepsilon}
  =a(\alpha,\beta).
\]
The limit is nonnegative and cannot vanish because the zero eigenvalue is simple. For any two positive values of $\alpha$ with $\beta$ fixed, the ordering of the regularized determinants passes to the limit. This proves monotonicity of $a$ in $\alpha$; fixing $\alpha$ gives monotonicity in $\beta$. Since $a$ is a polynomial, its two partial derivatives are nonnegative throughout the open positive quadrant.
\end{proof}

\paragraph{Dependence on the sum/maximum norm structure.}
The column-splicing and harmonic-extension steps in Lemma~\ref{lem:schur} use the recursive sum/maximum norm exactly. Common contractivity in an arbitrary norm is insufficient. For example,
\[
U=\begin{pmatrix}1/20&3/4\\39/40&1/20\end{pmatrix},\qquad
V=\begin{pmatrix}97/100&591/10000\\0&97/100\end{pmatrix}
\]
are entrywise nonnegative Euclidean contractions. To check their norms, the upper-left entry and determinant of $I-U^\top U$ are $3/64$ and $8289/640000$, respectively. For $I-V^\top V$ they are $591/10000$ and $0$. The first symmetric matrix is positive definite and the second is positive semidefinite, so $\|U\|_2<1$ and $\|V\|_2\le1$ for the induced Euclidean operator norm. Direct inversion gives $\tr((I-U)^{-1}(I-V))=-249/68500$.
Thus the treeplex proof uses the recursive $\ell_1/\ell_\infty$ structure established in Appendix~\ref{app:efg-prelims}.

%% file: appendices/fixed-point-stability.tex
\section{From Denominator Monotonicity to Fixed-Point Stability}\label{app:trace-stability}
\begin{proof}[Proof of Theorem~\ref{thm:efg-stability}]
When $m=1$, $w^\top s=1$ fixes the sole coordinate and strategy movement is zero. Henceforth let $m\ge2$. Fix $A>0$ in the lift and a nonzero matrix $V$ satisfying its homogeneous affine constraints. Set
\[
\eta=\max_{a,b}\frac{|V_{ab}|}{A_{ab}},\qquad
A^\pm=A\pm V/\eta.
\]
Both $A^+$ and $A^-$ are feasible and nonnegative.  For $\alpha,\beta>0$, define
\[
L(\alpha,\beta)=\alpha(I-A^+)+\beta(I-A^-),\qquad
\widetilde A(\alpha,\beta)=\frac{\alpha A^++\beta A^-}{\alpha+\beta}.
\]
The matrix $\widetilde A(\alpha,\beta)$ is positive.  We write $s_{\alpha,\beta}$ and $l_{\alpha,\beta}$ for its right and left Perron vectors, normalized by
\[
w^\top s_{\alpha,\beta}=1,\qquad l_{\alpha,\beta}^\top s_{\alpha,\beta}=1.
\]
For $s\in P$, each $\widetilde A^k s$ is feasible and has $w$-mass one. Perron--Frobenius convergence gives $\lim_{k\to\infty} \widetilde A^k s = s_{\alpha,\beta}(l_{\alpha,\beta}^\top s)$. Taking $w$-mass in this limit proves $l_{\alpha,\beta}^\top s=1$, in particular for $s=s^0$. This step uses the mass identity only on feasible strategies.

We set
\[
a(\alpha,\beta)=\tr(\adj L(\alpha,\beta)),\qquad
\mathbf u(\alpha,\beta)=\adj L(\alpha,\beta)s^0.
\]
Since $L(\alpha,\beta)=(\alpha+\beta)(I-\widetilde A(\alpha,\beta))$ has rank $m-1$, every column of $\adj L(\alpha,\beta)$ lies in $\ker L(\alpha,\beta)=\operatorname{span}\{s_{\alpha,\beta}\}$ and every row lies in the left kernel $\operatorname{span}\{l_{\alpha,\beta}^\top\}$.  Thus $\adj L(\alpha,\beta)=c\,s_{\alpha,\beta}l_{\alpha,\beta}^\top$ for some scalar $c$.  Taking traces and using $l_{\alpha,\beta}^\top s_{\alpha,\beta}=1$ gives $c=a(\alpha,\beta)$, so
\[
s_{\alpha,\beta}=\frac{\mathbf u(\alpha,\beta)}{a(\alpha,\beta)}.
\]
Each adjugate entry is a signed $(m-1)$-minor of a matrix linear and homogeneous in $(\alpha,\beta)$. Thus $a$ and each nonzero coordinate of $\mathbf u$ are homogeneous of degree $r=m-1$. Since $a>0$ on the positive quadrant, the quotient above is differentiable along every segment we use.  Lemma~\ref{lem:perron-denominator} gives
\[
\frac{\partial a(\alpha,\beta)}{\partial \alpha}\ge0,
\qquad
\frac{\partial a(\alpha,\beta)}{\partial \beta}\ge0.
\]

We write
\[
\rho(t)=a(1+t,1-t).
\]
Euler's identity, coordinatewise monotonicity, and homogeneity give the successive bounds
\begin{align*}
|\rho'(0)|
&\le \left.\frac{\partial a}{\partial \alpha}\right|_{(1,1)}+
      \left.\frac{\partial a}{\partial \beta}\right|_{(1,1)}
 =r\rho(0),\\
\rho(t)
&\le a(1+|t|,1+|t|)
 =(1+|t|)^r\rho(0)
 <2\rho(0),\qquad |t|\le\frac{1}{2r}.
\end{align*}
The first line uses Euler's identity together with nonnegative partial derivatives; the second uses monotonicity followed by homogeneity and the bound $(1+1/(2r))^r<2$.

Fix an affine test $P\to[0,1]$ as in Section~\ref{sec:model}. Subtracting $1/2$ preserves all differences on $P$. Since the mass functional satisfies $w^\top s=1$ there, the centered affine test can be represented by a vector $f$ with
\[
|f^\top s|\le\frac12
\qquad(s\in P).
\]
We define $F(t):=f^\top \mathbf u(1+t,1-t)$, a polynomial of degree at most $r$.  For $|t|\le 1/(2r)$, the bounds on $f$ and $\rho$ give
\[
|F(t)|=\rho(t)\,|f^\top s_{1+t,1-t}|\le\tfrac12\rho(t)<\rho(0).
\]
For every trigonometric polynomial $p$ of degree at most $r$, Bernstein's inequality \citep{queffelec2019} gives
\[
\max_{\theta\in\R}|p'(\theta)|\le r\max_{\theta\in\R}|p(\theta)|.
\]
Apply it to $G(\theta)=F(\sin\theta/(2r))$. Each power $(\sin\theta)^j$ has frequencies of absolute value at most $j$, so $G$ has trigonometric degree at most $r$. The bound on $F$ gives $\max_\theta|G(\theta)|\le\rho(0)$ since $|\sin\theta|\le1$. By the chain rule and Bernstein's inequality,
\[
\frac{|F'(0)|}{2r}=|G'(0)|\le r\max_\theta|G(\theta)|\le r\rho(0).
\]
Thus $|F'(0)|\le2r^2\rho(0)$. Combining this with $|\rho'(0)|\le r\rho(0)$ and $|F(0)|\le \rho(0)/2$ yields
\[
\left|\left(\frac{F}{\rho}\right)'(0)\right|
\le\frac{|F'(0)|}{\rho(0)}+\frac{|F(0)|\,|\rho'(0)|}{\rho(0)^2}
\le2r^2+\frac r2
\le\frac52r^2.
\]
Since
\[
\widetilde A(1+t,1-t)=A+tV/\eta,
\]
rescaling the derivative multiplies the preceding bound by $\eta$. The derivative lies in the direction space of $P$, since the fixed point stays in $P$ along the line. Taking the supremum over the centered affine tests therefore gives the local sensitivity bound
\begin{equation}
\left\|\left.\frac{d}{dt}s_{A+tV}\right|_{t=0}\right\|_P
\le\frac52(m-1)^2\max_{a,b}\frac{|V_{ab}|}{A_{ab}}
\le\frac52(m-1)^2
\left(\sum_{a,b}\frac{V_{ab}^2}{A_{ab}^2}\right)^{1/2}.
\label{eq:local-efg}
\end{equation}

It remains to integrate along $A_t=(1-t)A+tB$ for $0\le t\le1$. The case $A=B$ is immediate. If $D_R(B,A)\ge1/16$, then $\norm{s_B-s_A}_P^2\le1\le16D_R(B,A)$. Otherwise, let $g(u)=u-1-\log u$. Each summand $g(B_{ab}/A_{ab})$ in the divergence is nonnegative. The function decreases up to $1$ and increases thereafter, and $g(1/2),g(2)>1/16$. Hence every ratio lies in $(1/2,2)$. Write $u_{ab}:=B_{ab}/A_{ab}$. Since $R(A)=-\sum_{a,b}\log A_{ab}$,
its gradient has entries $(\nabla R(A))_{ab}=-1/A_{ab}$. Thus
\begin{align}
D_R(B,A)
&=R(B)-R(A)-\langle\nabla R(A),B-A\rangle \notag\\
&=\sum_{a,b}\left(-\log B_{ab}+\log A_{ab}
                  +\frac{B_{ab}-A_{ab}}{A_{ab}}\right) \notag\\
&=\sum_{a,b}\bigl(u_{ab}-1-\log u_{ab}\bigr)
 =\sum_{a,b}g(u_{ab}). \notag
\end{align}
For $u\in[1/2,2]$, Taylor's theorem about $1$, together with
$g(1)=g'(1)=0$, gives $g(u)=\tfrac12 g''(\xi)(u-1)^2$ for some
$\xi$ between $1$ and $u$. Since $g''(\xi)=1/\xi^2\ge 1/4$ on this
interval, $g(u)\ge (u-1)^2/8$. Applying this inequality to each
$u_{ab}$ and summing gives
\begin{align}
\sum_{a,b}(u_{ab}-1)^2
&\le 8\sum_{a,b}g(u_{ab})=8D_R(B,A). \notag
\end{align}

We next bound the relative change of each entry along
$A_t=(1-t)A+tB$. For every $t\in[0,1]$, the bound $u_{ab}\ge 1/2$
gives
\begin{align}
(A_t)_{ab}
&=A_{ab}(1-t+tu_{ab})
 \ge A_{ab}(1-t/2)
 \ge A_{ab}/2. \notag
\end{align}
It follows that, uniformly in $t$,
\begin{align}
\sum_{a,b}\frac{(B_{ab}-A_{ab})^2}{(A_t)_{ab}^2}
&\le 4\sum_{a,b}\frac{(B_{ab}-A_{ab})^2}{A_{ab}^2} \notag\\
&=4\sum_{a,b}(u_{ab}-1)^2
 \le 32D_R(B,A). \notag
\end{align}

Each $A_t$ is positive and feasible, and the segment has direction
$B-A$. Applying~\eqref{eq:local-efg} at $A_t$ in this direction gives
\begin{align}
\left\|\frac{d}{dt}s_{A_t}\right\|_P
&\le \tfrac52(m-1)^2
     \left(\sum_{a,b}\frac{(B_{ab}-A_{ab})^2}{(A_t)_{ab}^2}\right)^{1/2}
     \notag\\
&\le \tfrac52(m-1)^2\sqrt{32D_R(B,A)}. \notag
\end{align}
Since $A_0=A$ and $A_1=B$, the fundamental theorem of calculus and the
triangle inequality now give
\begin{align}
\norm{s_B-s_A}_P
&=\left\|\int_0^1 \frac{d}{dt}s_{A_t}\,dt\right\|_P \notag\\
&\le \int_0^1\left\|\frac{d}{dt}s_{A_t}\right\|_P\,dt
 \le \tfrac52(m-1)^2\sqrt{32D_R(B,A)}. \notag
\end{align}
Squaring both sides and using $(5/2)^2\cdot 32=200$ yields
\begin{align}
\norm{s_B-s_A}_P^2
&\le 200(m-1)^4D_R(B,A)
 \le 200m^4D_R(B,A). \notag
\end{align}
\end{proof}

%% file: appendices/self-play-regret.tex
\section{From Fixed-Point Stability to Self-Play Regret}\label{app:closure}
\subsection{Equivalence of the score and proximal forms}
Fix a player and suppress its index. The affine hull of $\cZ$ is given by the lift equalities. The entry-log barrier tends to infinity at its relative boundary and has positive-definite Hessian on tangent directions. Thus every barrier-plus-linear objective in Algorithm~\ref{alg:exact} has a unique positive minimizer. At $\bar A^t$ its directional derivative vanishes on the whole tangent space. Hence on $\cZ$ the function
\[
A\longmapsto \gamma R(A)+\sum_{\tau<t}c^\tau(A)-\gamma D_R(A,\bar A^t)
\]
is constant: its linear part is $\gamma\nabla R(\bar A^t)+\sum_{\tau<t}c^\tau$, which annihilates feasible differences. Adding either the prediction or the realized cost therefore gives
\begin{align*}
A^t&=\arg\min_{A\in\cZ,\,A>0}
 \{c^{t-1}(A)+\gamma D_R(A,\bar A^t)\},\\
\bar A^{t+1}&=\arg\min_{A\in\cZ,\,A>0}
 \{c^t(A)+\gamma D_R(A,\bar A^t)\}.
\end{align*}
Both updates start from $\bar A^t$. The first determines play before observing round $t$'s loss. The second is a leader used for analysis after that loss is observed. At $t=1$, $\bar A^1=A^1=A_c$ since $c^0=0$.

\subsection{Directed divergences and strong convexity}
For a linear cost $g$ and positive $U$, let $Y$ minimize $g+\gamma D_R(\cdot,U)$. For any positive comparator $X$, first-order optimality along $X-Y$ and the definition of $D_R$ give
\[
g(Y)-g(X)=\gamma[D_R(X,U)-D_R(X,Y)-D_R(Y,U)].
\]
Apply this identity first with $(g,U,Y,X)=(c^t,\bar A^t,\bar A^{t+1},C)$ and then with $(c^{t-1},\bar A^t,A^t,\bar A^{t+1})$. Add the two identities and add the difference between the realized and predicted costs on $A^t-\bar A^{t+1}$. The $D_R(\bar A^{t+1},\bar A^t)$ terms cancel, leaving
\begin{align*}
c^t(A^t)-c^t(C)
={}&\gamma[D_R(C,\bar A^t)-D_R(C,\bar A^{t+1})]\\
&-\gamma D_R(A^t,\bar A^t)-\gamma D_R(\bar A^{t+1},A^t)\\
&+(c^t-c^{t-1})(A^t-\bar A^{t+1}).
\end{align*}
Here a linear cost on a matrix difference denotes the difference of the two costs.

For this proof write $\|V\|_{\cZ-\cZ}:=\inf\{r>0:V\in r(\cZ-\cZ)\}$ on the space of feasible differences. This is the same difference-body norm used for strategies in Appendix~\ref{app:polyhedral}, now on the lift. If $V\ne0$ is tangent at a positive $A$, set
\[
h_A(V):=\left(\sum_{a,b} V_{ab}^2/A_{ab}^2\right)^{1/2}.
\]
Both $A\pm V/h_A(V)$ satisfy the lift equalities and nonnegativity. Their difference is $2V/h_A(V)$, so $2\|V\|_{\cZ-\cZ}\le h_A(V)$. The same inequality holds trivially for $V=0$. Setting $V=B-A$ for positive $A,B$ and using the integral remainder formula gives
\[
D_R(B,A)=\int_0^1(1-u)\,h_{A+uV}(V)^2\,du
 \ge 2\|B-A\|_{\cZ-\cZ}^2.
\]
In particular the weaker bound $\|B-A\|_{\cZ-\cZ}\le\sqrt{2D_R(B,A)}$ suffices below. The definition of $\Delta^t$ implies $|(c^t-c^{t-1})(U-V)|\le\Delta^t$ for $U,V\in\cZ$. By scaling this gives $|(c^t-c^{t-1})(H)|\le\Delta^t\|H\|_{\cZ-\cZ}$ on the tangent space. Apply it with $H=A^t-\bar A^{t+1}$ and use Young's inequality:
\[
|(c^t-c^{t-1})(H)|
\le\Delta^t\sqrt{2D_R(\bar A^{t+1},A^t)}
\le\frac{(\Delta^t)^2}{\gamma}+\frac{\gamma}{2}D_R(\bar A^{t+1},A^t).
\]
Summing in $t$ telescopes the comparator divergences. The final term $-\gamma D_R(C,\bar A^{T+1})$ is nonpositive. Each of the two movement divergences retains coefficient at most $-\gamma/2$, which proves~\eqref{eq:rvu}.

\subsection{Loss variation and the common rate theorem}
Restore the player index. For any two comparator maps $B,C$ and $t\ge2$, linearity of the deviation costs gives
\begin{align*}
&(c_i^t-c_i^{t-1})(B-C)\\
&=(1-\lambda)\Bigl[\ip{\ell_i^t-\ell_i^{t-1}}{(B-C)x_i^t}
+\ip{\ell_i^{t-1}}{(B-C)(x_i^t-x_i^{t-1})}\Bigr].
\end{align*}
The vector $(B-C)x_i^t$ is a difference between two feasible strategies, so its strategy norm is at most one. Write $v_j^t=\|x_j^t-x_j^{t-1}\|_j$. Since $B$ and $C$ are nonexpansive and $\|\ell_i^{t-1}\|_{i,*}\le2$, the second inner product has absolute value at most $4v_i^t$. By~\eqref{eq:smoothness}, squaring the sum and using $(a+b)^2\le2a^2+2b^2$ gives
\[
(\Delta_i^t)^2\le32(n-1)\sum_{j\ne i}(v_j^t)^2+32(v_i^t)^2.
\]
When we sum over $i$, each $(v_j^t)^2$ occurs $n-1$ times in the opponent sums. At $t=1$, $c_i^0=0$ and $\Delta_i^1\le2$. Thus
\[
\sum_{i,t}(\Delta_i^t)^2\le4n+32[1+(n-1)^2]\sum_iV_i
\le4n+32n^2\sum_iV_i.
\]
For the second part of Lemma~\ref{lem:variation-main}, insert the fixed point of $\bar A_i^t$ between $x_i^{t-1}$ and $x_i^t$. The squared triangle inequality and the assumed directed stability give
\[
(v_i^t)^2\le2C_*\bigl[D_{R_i}(\bar A_i^t,A_i^{t-1})+
D_{R_i}(A_i^t,\bar A_i^t)\bigr].
\]
The first divergence is the correction term at $t-1$, the second the prediction term at $t$. Summing from $2$ through $T$ uses each of these terms at most once and gives $V_i\le2C_*E_i$.

To prove Theorem~\ref{thm:exact}, fix any comparator $C\in\cZ_i$. Since $x_i^t=(1-\lambda)A_i^t x_i^t+\lambda q_i^0$,
\[
\ip{\ell_i^t}{x_i^t-Cx_i^t}
=c_i^t(A_i^t)-c_i^t(C)+\lambda\ip{\ell_i^t}{q_i^0-Cx_i^t}.
\]
The last inner product is bounded above by two. To allow a comparator on the boundary, set $\eps=1/(T+1)$ and $C_\eps=(1-\eps)C+\eps A_{c,i}$. Positivity and stationarity of $A_{c,i}$ yield
\[
D_{R_i}(C_\eps,A_{c,i})=R_i(C_\eps)-R_i(A_{c,i})
\le m_i^2\log(1/\eps),
\]
since $(C_\eps)_{ab}\ge\eps(A_{c,i})_{ab}$ for each entry. Also $|c_i^t(C_\eps)-c_i^t(C)|\le2(1-\lambda)\eps$, so the entire smoothing cost is at most two. Apply~\eqref{eq:rvu} to $C_\eps$ and use the preceding reset identity. These estimates are uniform in $C$. Proposition~\ref{prop:slack} lifts a minimizing hindsight deviation of each player, so summing the corresponding inequalities gives
\[
\sum_i\LinReg_i(T)\le
\gamma\Bigl(\sum_i m_i^2\Bigr)\log(T+1)+2n(\lambda T+1)
+\frac{4n}{\gamma}
+\left(\frac{64n^2 C_*}{\gamma}-\frac{\gamma}{2}\right)\sum_iE_i.
\]
The movement coefficient is nonpositive if $\gamma^2\ge128n^2 C_*$, proving~\eqref{eq:master}. Choosing the identity as a candidate in each player's hindsight minimization shows that the minimum is at most that player's incurred loss. Thus each regret is nonnegative, and the sum bound also bounds each individual regret.

For the general-polytope schedule, $a_T=\lceil(T+1)^{1/3}\rceil$, $\lambda=a_T^{-2}$ and $C_*\le2(p+1)^2a_T^2$. Substitution gives
\[
\sum_i\LinReg_i(T)\le
 32n(p+1)a_T\Bigl(\sum_i m_i^2\Bigr)\log(T+1)
 +\frac{1}{8(p+1)a_T}+2n(T/a_T^2+1).
\]
Both $a_T$ and $T/a_T^2$ are $\mathcal{O}((T+1)^{1/3})$, giving the claimed horizon dependence. In EFG mode $\lambda=0$, $C_*=200M^4$ and $\gamma=320nM^2$. Using $\sum_i m_i^2\le nM^2$ gives $320n^2M^4\log(T+1)+1/(80M^2)+2n$ for the matrix lift.

%% file: appendices/finite-precision.tex
\section{Deterministic Finite-Precision Implementation}\label{app:implementation}
This section proves the bit statement for the explicit logarithmic learner in both modes. The computational primitives are weak convex optimization over the rational positive lift and rational linear algebra for the fixed-point system. Input polytopes and the game description are rational. A full-information feedback procedure returns own loss coefficients to requested polynomial-bit precision; in an explicit rational EFG these coefficients are obtained by summing terminal contributions at the current rational profile.

\subsection{Positive coordinate floors and deterministic optimization}
We fix a rational positive feasible anchor $B_0$. At a minimizer $A$ of $\gamma R+b$, optimality along $B_0-A$ gives
\[
\gamma\left(m_i^2-\sum_{ab}\frac{(B_0)_{ab}}{A_{ab}}\right)+b(B_0-A)=0.
\]
Thus every coordinate satisfies
\[
A_{ab}\ge\frac{(B_0)_{ab}}{m_i^2+(\max_{\cZ_i}b-\min_{\cZ_i}b)/\gamma}.
\]
Through horizon $T$, $\max_{\cZ_i}b-\min_{\cZ_i}b\le2(T+1)$ because each score sums at most $T+1$ single-round losses and each can differ by at most two between feasible matrices. When scores are rounded, we increase this bound by a conservative unit. We restrict optimization to entries at least half the resulting floor. This rational truncation contains the relevant exact minimizer and has a relative interior ball of polynomially encoded radius around $B_0$. The affine equalities and coordinate upper bounds also have polynomial bit length. On this domain the barrier gradient is a rational reciprocal vector, and logarithmic objective values have deterministic rational enclosures to any requested precision. The magnitudes of derivatives and the ratio of encoded outer to inner radii may be large, but their logarithms are polynomial in $D$ and $\log(T+1)$.

Deterministic weak convex optimization with a separation oracle therefore finds a positive rational feasible point to any prescribed inverse-polynomial objective gap in $\poly(D,\log(T+1))$ bit time \citep{gls1988}. Computation takes place in a rational basis of the affine hull; exact feasibility can be preserved by rounding toward the interior anchor before coordinate rounding. This optimization step depends only on the rational lift geometry and the requested objective accuracy.

\subsection{Regret of the finite-precision iterates}
We denote the common squared-stability bound by $C_*$, including its dependence on $T$ in the general case. Both choices~\eqref{eq:schedules-general} and~\eqref{eq:schedules-efg} satisfy $\gamma^2\ge512n^2C_*$. We set
\begin{equation}
\delta_T=\frac1{4096n^2(1+C_*)(T+1)^2}.
\label{eq:precision-budget}
\end{equation}
We write $\widehat A_i^t\in\cZ_i$ for the positive feasible matrix actually returned by the finite-precision optimization routine, and we let
$\widehat x_i^t:=x_{\widehat A_i^t}^\lambda$ be its exact fixed point.  For the analysis, we define the exact prediction leaders $A_i^t$ and correction leaders $\bar A_i^t$ from the true costs generated at these \emph{actual implemented profiles}:
\[
c_i^t(A)=(1-\lambda)\ip{\ell_i(\widehat x_{-i}^t)}{A\widehat x_i^t}.
\]
Thus $A_i^t$ and $\bar A_i^t$ are the exact counterparts of the prediction and correction updates in Section~\ref{sec:analysis}, evaluated on the implemented feedback sequence.  We write
\[
x_i^{*,t}:=x_{A_i^t}^\lambda,
\qquad
\widehat V_i:=\sum_{t=2}^T\norm{\widehat x_i^t-\widehat x_i^{t-1}}_i^2,
\qquad
V_i^*:=\sum_{t=2}^T\norm{x_i^{*,t}-x_i^{*,t-1}}_i^2,
\]
and let $E_i$ denote the barrier energy of the exact shadow leaders, as defined in Section~\ref{sec:analysis}.  The shadow leaders compare the implemented iterates with the exact prox solutions.  For either the prediction or correction prox, we write $b$ for the exact linear score and $\widehat b$ for its rounded implementation.  For the score error $e:=\widehat b-b$, we require
\[
\max_{\cZ_i}e-\min_{\cZ_i}e\le\frac{\gamma\delta_T}{2},
\qquad
(\gamma R_i+\widehat b)(\widehat A)-\min_{A\in\cZ_i,\,A>0}(\gamma R_i+\widehat b)(A)\le\frac{\gamma\delta_T}{2}.
\]
If $A^*$ denotes the exact minimizer for the score $b$, we combine these two bounds as
\begin{align*}
(\gamma R_i+b)(\widehat A)-(\gamma R_i+b)(A^*)
&\le\max_{\cZ_i}e-\min_{\cZ_i}e+\frac{\gamma\delta_T}{2}\\
&\le\gamma\delta_T.
\end{align*}
The first line separates score-rounding and optimization errors; the second uses the prescribed half-budget for each.
The first-order condition at the positive minimizer $A^*$ makes the left-hand side exactly $\gamma D_{R_i}(\widehat A,A^*)$.  Applied to the prediction leader, this gives
\[
D_{R_i}(\widehat A_i^t,A_i^t)\le\delta_T.
\]
The rounded matrix remains positive and feasible. Its reset fixed point is found by rational linear algebra. For $\lambda=0$, we solve the rational nullspace equation with mass normalization; the Perron eigenvalue is simple. For $\lambda>0$, the contracting linear system is nonsingular. Determinant bit bounds give a rational solution of polynomial encoding length, even if intermediate numerical condition numbers are large.

Stability gives $\norm{\widehat x_i^t-x_i^{*,t}}_i^2\le C_*\delta_T$. Thus the actual path length obeys
\[
\widehat V_i\le3V_i^*+6TC_*\delta_T,
\qquad V_i^*\le2C_*E_i.
\]
The actual-profile variation bound now gives an energy coefficient at most
\[
\frac{192n^2C_*}{\gamma}-\frac\gamma2\le-\frac\gamma8.
\]
The additional variation cost is at most $192n^3TC_*\delta_T/\gamma$.  The strong-convexity estimate above and $D_{R_i}(\widehat A_i^t,A_i^t)\le\delta_T$ imply
\[
\inf\{r>0:\widehat A_i^t-A_i^t\in r(\cZ_i-\cZ_i)\}
\le\sqrt{2\delta_T}.
\]
Hence the approximate Gordon identity contributes at most
\[
\sum_{i=1}^n\sum_{t=1}^T2\inf\{r>0:\widehat A_i^t-A_i^t\in r(\cZ_i-\cZ_i)\}
\le2nT\sqrt{2\delta_T}.
\]
Indeed, the implemented strategy is an exact fixed point of its implemented map, so the discrepancy is $(1-\lambda)\ell_i^{t\top}(\widehat A_i^t-A_i^t)\widehat x_i^t$. We combine the two errors as
\[
2nT\sqrt{2\delta_T}
+\frac{192n^3TC_*\delta_T}{\gamma}<1,
\]
where the strict bound follows from~\eqref{eq:precision-budget}, $C_*\ge1$, and $\gamma\ge n$. The comparator and reset terms in~\eqref{eq:master} are unchanged.

Finally, score rounding and approximate feedback can meet the stated error budget. Compactness and Appendix~\ref{app:polyhedral} give a rational encoded bound on every matrix coordinate. If each cost coefficient is perturbed by at most $\xi$, the resulting error in any cost difference between two feasible matrices is at most
\[
2m_i^2 \xi\,\max_{A\in\cZ_i}\max_{a,b}A_{ab}.
\]
Taking per-round errors inverse-polynomial in $T$, $C_*$, and this encoded bound gives the required total budget. Their bit lengths are polynomial in $D,\log(T+1)$, because $\lambda^{-1}$ and $C_*$ are polynomial in $T$ and the compact description. This proves the deterministic implementation claim and its realized-regret guarantee.

%% file: appendices/unknown-horizons.tex
\section{Unknown Horizons and the Generality Boundary}\label{app:anytime}
For the general-polytope mode, common epochs of intended lengths $1,2,4,\ldots$ preserve the $\widetilde{\mathcal{O}}(\poly(D)T^{1/3})$ rate. The bound for a horizon-$H$ instance also applies to every prefix, since smoothing, parameters, and error budgets were chosen for $H$. For any fixed linear deviation, cumulative regret is at most the sum of epochwise comparator maxima. If $s$ is the last active epoch, the geometric sums $\sum_{k\le s}2^{k/3}=\mathcal{O}(T^{1/3})$ and $\sum_{k\le s}(k+1)2^{k/3}=\mathcal{O}(T^{1/3}\log(T+1))$ prove the claim.

The pure-log EFG algorithm uses a horizon-independent $\gamma$ and no reset, so its exact logarithmic theorem holds without restarts. Its stated finite-bit implementation is horizon-specified. An anytime finite-bit implementation can use summable time-dependent accuracy budgets while reoptimizing the cumulative rational scores at each round.

The generic theorem applies to arbitrary compact $H$-polytopes with polynomial-size descriptions, while the EFG theorem applies to arbitrary-depth perfect-recall treeplexes, including imperfect-information games. The sharper EFG rate is driven by the sum/maximum norm identity behind unreset stability. For a general $H$-polytope, Lemma~\ref{lem:comparison} instead transfers the regenerative estimate to the explicit barrier, retaining the inverse-reset cost and the resulting $T^{1/3}$ balance.

%% file: appendices/utc-implementation.tex
\section{UTC Implementation of the EFG Algorithm}\label{app:utc-implementation}
This appendix proves Corollary~\ref{cor:utc-implementation}. We give the play and query flow constraints, show that the induced maps cover all linear deviations, and transfer the local treeplex stability estimate to the UTC log barrier. This yields the coordinate update used in Appendix~\ref{app:experiments}.

\subsection{Stability under a nonnegative coordinate representation}
The transfer uses the local relative-perturbation estimate already proved in~\eqref{eq:local-efg}; it does not compare two different global matrix-barrier divergences.

\begin{lemma}[Stability under a nonnegative linear representation]\label{lem:utc-lift-inheritance}
We consider a treeplex $P$ with $m$ coordinates indexed by terminal sequences and a compact polytope $Z=\{z\in\R_+^q:Bz=b_{\rm U}\}$ containing a strictly positive point.  The linear map $C:\R^q\to\R^{m\times m}$ satisfies $C(Z)\subseteq\cZ(P)$, every entry of $C(z)$ is a nonnegative linear combination of the coordinates of $z$, and $C(z)>0$ whenever $z>0$.  With $\widehat R(z):=-\sum_{e=1}^q\log z_e$, every positive $z,z'\in Z$ satisfies
\[
\norm{s_{C(z')}-s_{C(z)}}_P^2
\le200m^4D_{\widehat R}(z',z).
\]
If the maps $s\mapsto C(z)s$, $z\in Z$, cover all affine self-maps of $P$, then the common regret analysis applies on $Z$, with $q$ in place of $m^2$ in the comparator-radius term.
\end{lemma}
\begin{proof}
The singleton case is immediate, so we assume $m\ge2$.  We write $C(z)_{ab}=\sum_{e=1}^q c_{ab,e}z_e$, where all coefficients $c_{ab,e}$ are nonnegative.  For any feasible direction $h$ and positive $z$, we combine the relative-coordinate estimate with~\eqref{eq:local-efg} as
\begin{align*}
\frac{\left|\sum_{e=1}^q c_{ab,e}h_e\right|}{C(z)_{ab}}
&\le\frac{\sum_{e=1}^q c_{ab,e}|h_e|}{\sum_{e=1}^q c_{ab,e}z_e}
\le\max_{1\le e\le q}\frac{|h_e|}{z_e},\\
\left\|\left.\frac{d}{dt}s_{C(z+th)}\right|_{t=0}\right\|_P
&\le\frac52(m-1)^2\max_{1\le e\le q}\frac{|h_e|}{z_e}.
\end{align*}
The first line uses nonnegativity of the coefficients $c_{ab,e}$, and the second is the local matrix estimate applied to $C(z)$ and $C(h)$.

If $D_{\widehat R}(z',z)\ge1/16$, the intrinsic diameter bound proves the claim.  Otherwise every ratio $z'_e/z_e$ lies in $(1/2,2)$, and along the segment we have
\begin{align*}
D_{\widehat R}(z',z)
&=\sum_{e=1}^q\left(\frac{z'_e}{z_e}-1-\log\frac{z'_e}{z_e}\right),\\
\sum_{e=1}^q\left(\frac{z'_e}{z_e}-1\right)^2
&\le8D_{\widehat R}(z',z),\\
\max_{1\le e\le q}\frac{|z'_e-z_e|}{(1-t)z_e+tz'_e}
&\le2\left(\sum_{e=1}^q\left(\frac{z'_e}{z_e}-1\right)^2\right)^{1/2}
\le2\sqrt{8D_{\widehat R}(z',z)},\\
\norm{s_{C(z')}-s_{C(z)}}_P^2
&\le200(m-1)^4D_{\widehat R}(z',z)\le200m^4D_{\widehat R}(z',z).
\end{align*}
The first line is the coordinate formula for the UTC entry-log barrier.  The second again uses the curvature of $u-1-\log u$ on $(1/2,2)$, and the third uses the segment lower bound $(1-t)z_e+tz'_e\ge z_e/2$.  The last line follows by integrating the preceding derivative estimate and squaring.  Under the surjectivity hypothesis in the last sentence of the lemma, every affine self-map has a representative in $Z$. Since the cost depends only on the induced map, maximizing over two points of $Z$ gives exactly the same loss gap in~\eqref{eq:prediction-gap} as maximizing over two linear deviations.  The Gordon identity, log-barrier strong convexity, and analytic-center comparator smoothing apply to the positive affine slice $Z$ just as in Appendix~\ref{app:closure}.  Its barrier contains $q$ logarithmic terms, which gives the stated comparator-radius substitution.
\end{proof}

\subsection{The UTC flow constraints and their induced maps}
Suppress the player index. Write $A(z)$ for the sequence-pair matrix in the UTC formulation and $H(z)$ for its auxiliary matrix indexed by pairs of information sets \citep[Eq.~(2)]{zhang2024}. Only $A(z)$ is applied to a strategy. Its root row is fixed by $A_{\varnothing,\sigma}=\mathbf 1_{\{\sigma=\varnothing\}}$. The vector $z$ consists of every other entry of $A$ and every entry of $H$, all nonnegative. For $\sigma\ne\varnothing$ let $j(\sigma)$ be the information set at which the last action of $\sigma$ was taken. Write $\mathcal C(\sigma)=\{k:p(k)=\sigma\}$ and interpret $H_{j,j(\varnothing)}$ as zero. The flow equation at $(j,\sigma)$ is
\[
 A_{p(j),\sigma}+H_{j,j(\sigma)}
 =\sum_{a\in\mathcal A_j} A_{ja,\sigma}
   +\sum_{k\in\mathcal C(\sigma)}H_{j,k}.
\]
Its right-hand side lists the moves available to the deviator: play an action at $j$ or query the mediator at a successor $k$ of the current query sequence $\sigma$. In matrix form these equations are $Bz=b_{\rm U}$ after substituting the root row. Thus the number of coordinates entering the barrier is
\[
q=(|\Sigma|-1)|\Sigma|+|\mathcal J|^2.
\]
With the flow equalities imposed, the affine dimension is $q-\operatorname{rank}B$ once a positive feasible point exists.

To construct such a point, visit the pairs in increasing total depth of the play and query sequences and divide the incoming mass equally among all entries on the right. Each output entry is assigned at exactly one pair. At $(j,\varnothing)$ one can reach $j$ from a root information set using play actions alone. Keeping $j$ fixed, one can next reach $(j,\sigma)$ by making the queries along $\sigma$. The initial incoming mass is one, so each such path carries positive mass. Every pair is therefore reached and every stored coordinate is positive. The same depth ordering gives compactness: at a pair every output is at most the sum of its at most two previously bounded inputs. The resulting closed rational polyhedron is bounded. A rational positive anchor of polynomial encoding length can equivalently be found by maximizing a common lower bound on its coordinates by linear programming.

Let $\mathsf S$ select the terminal sequences and let $\mathsf T$ be the nonnegative reconstruction from Appendix~\ref{app:efg-prelims}. In particular its rows on terminal sequences form the identity. Then
\[
\mathsf S\mathsf T=I,\qquad \mathsf T\mathsf S x=x\quad(x\in Q),
\qquad C(z)=\mathsf S A(z)\mathsf T.
\]
By \citet[Thm.~4.1]{zhang2024}, the maps $x\mapsto A(z)x$ on $Q$ comprise all linear deviations. Hence $C(z)$ maps $P$ into itself, preserves its affine hull, and represents every affine self-map of $P$. For a player with a decision, each terminal output row is a nonroot row of $A$. Thus each entry of $C(z)$ is a nonnegative linear combination of stored coordinates. For terminal $a,b$, the identity rows of $\mathsf T$ give $C(z)_{ab}\ge A(z)_{a,b}>0$ whenever $z>0$. Players with singleton strategy sets have zero regret. This proves all hypotheses of Lemma~\ref{lem:utc-lift-inheritance} for the players being updated.

If player $i$ has $q_i$ stored UTC coordinates, the exact UTC updates therefore satisfy
\[
\sum_{i=1}^n\LinReg_i(T)
\le\gamma\left(\sum_{i=1}^n q_i\right)\log(T+1)
+\frac{4n}{\gamma}+2n,\qquad \gamma^2\ge128n^2(200M^4).
\]
The terminal-sequence matrix lift already gives the $M^4$ bound in Theorem~\ref{thm:informal}. For the UTC bound in terms of $M$ alone, first suppress one-action information sets: their sole child sequence has the same weight as its parent, so identifying those coordinates and reattaching subsequent information sets gives an affine isomorphism and can only decrease the number of terminal sequences. In the resulting tree on own sequences, connect each $ja$ to $p(j)$. Each nonleaf has at least two children, so $|\Sigma_i|\le 2m_i-1$. Furthermore, $|\Sigma_i|-1=\sum_{j\in\mathcal J_i}|\mathcal A_j|\ge2|\mathcal J_i|$. Hence $q_i\le5m_i^2$, giving $\mathcal O(n^2M^4\log(T+1))$ for the UTC update with the schedule in~\eqref{eq:schedules-efg}. With forced moves retained, the displayed bound in $q_i$ applies directly.

\subsection{Exact UTC-coordinate update}
Let $s^t\in P$ denote the strategy in terminal coordinates and $x^t=\mathsf T s^t\in Q$ its full realization plan. Write $\ell^t$ for the loss vector in terminal coordinates. If the feedback is a full-coordinate vector $g^t$, take $\ell^t=\mathsf T^\top g^t$. Set $c^0=0$ and let $c^t$ be the coefficient vector of the linear cost $z\mapsto\langle\ell^t,C(z)s^t\rangle$. Using the full-coordinate cost $\langle g^t,A(z)x^t\rangle$ gives the same value on feasible $z$, since $\mathsf T C(z)s^t=A(z)x^t$. The UTC update is
\[
 z^t=\arg\min_{z\in Z,\,z>0}
 \left\{\gamma\widehat R(z)+
 \left\langle\sum_{\tau=1}^{t-1}c^\tau+c^{t-1},z\right\rangle\right\},
 \qquad (I-C(z^t))s^t=0,\quad w^\top s^t=1.
\]
Positivity of $C(z^t)$ and Proposition~\ref{prop:slack} give a unique normalized solution in $P$; its reconstructed plan is fixed by $A(z^t)$. One may also solve the full system with $x_\varnothing^t=1$, as in the experiments. Indeed, write the nonroot blocks of $A$ as $[v\ G]$. Its positive feasible fixed plan, obtained by reconstruction, has nonroot part $y>0$ and satisfies $Gy=y-v<y$ because $v>0$. In the weighted maximum norm $\|u\|=\max_k|u_k|/y_k$, the nonnegative matrix $G$ has operator norm at most $\max_k(Gy)_k/y_k<1$. Therefore $(I-G)y=v$ has a unique solution. Both implementations require one barrier optimization and one rational linear solve for a rational feasible iterate. The finite-precision proof of Appendix~\ref{app:implementation} applies to $z$ with $q$ logarithmic terms and the stability estimate just proved.

%% file: appendices/experiments.tex
\section{Experimental Protocol and Results}\label{app:experiments}\label{sec:experiments}
We implement the UTC-coordinate EFG specialization of Algorithm~\ref{alg:exact} from Corollary~\ref{cor:utc-implementation} on four-player Kuhn, five-player Kuhn, and three-player Leduc, with $81$, $193$, and $457$ sequence coordinates per player and unreduced pure-plan counts of order $10^{12}$, $10^{29}$, and $10^{68}$.  At matched horizons under the same hindsight oracle for the full linear/UTC class, ORBITS lowers regret over UTC-CFR/RM+ by $17.1\%$, $25.4\%$, and $8.4\%$.  This appendix reports endpoints, protocol, trajectories, diagnostics, and parameters.

The experiments instantiate the UTC-coordinate EFG algorithm established in Appendix~\ref{app:utc-implementation}. This appendix specifies the benchmark instances, numerical solution of the theoretical update, evaluation protocol, diagnostics, and trajectories.

\subsection{Benchmarks and numerical implementation}
The experiments use four-player five-card Kuhn, five-player six-card Kuhn, and three-player one-bet Leduc.  Kuhn and Leduc are standard benchmark families \citep{kuhn1950,southey2005,zhang2024}; the five-player Kuhn instance is a larger member of the same family.  The games contain $3{,}960$, $58{,}320$, and $4{,}500$ terminal histories in the benchmark constructors, with suit-equivalent chance outcomes aggregated in Leduc. The losses are negative net chip payoffs divided by $6$, $8$, and $14$, respectively.  Each player has respectively $81$, $193$, and $457$ sequence coordinates, including the root. The respective intrinsic dimensions are $40$, $96$, and $228$. Each information set has two actions and contributes one independent strategy parameter in the relative interior. This gives unreduced pure-plan counts $2^{40}$, $2^{96}$, and $2^{228}$.  The UTC barrier-coordinate counts are $8{,}080$, $46{,}272$, and $260{,}376$ per player.  The corresponding numbers of UTC flow equalities are $3{,}240$, $18{,}528$, and $104{,}196$.

\paragraph{Numerical implementation of Algorithm~\ref{alg:exact}.}
The initial UTC point is the analytic center of the positive domain, computed by a separate zero-score solve.  Each subsequent round numerically solves the exact cumulative-score barrier problem stated in Appendix~\ref{app:utc-implementation}, then computes the corresponding unreset fixed point.  The accumulated score is always formed from the realized losses; warm starts affect only the numerical solver and do not change the update being implemented.

\paragraph{Numerical optimization and feasibility.}
For a cumulative score $b$, the scaled objective is $-\sum_e\log z_e+\langle b/\gamma,z\rangle$.  If $\eta$ is its (scaled-objective) equality multiplier, the dual slack is
\[
 s=b/\gamma+B^\top\eta>0,
 \qquad z_e=1/s_e.
\]
Damped Newton steps solve the dual equations $Bz=b_{\rm U}$.  Linear systems use sparse factorization for four-player Kuhn and preconditioned conjugate gradients for the larger instances.  Dual potentials obtained by the UTC backward recursion shift $b$ only by a vector in the row space of $B$, so the primal objective changes only by a constant.  The preceding dual solution supplies a warm start, but each objective is formed from the accumulated observed costs.

Before play, outgoing weights at each UTC node are converted to positive proportions and incoming mass is propagated forward through the UTC graph.  This produces a positive feasible point up to floating-point roundoff.  Both the strategy constraints and the fixed-point residual are then checked without silently clipping strategy coordinates.  Runs terminate with an error if a numerical solve fails its specified diagnostics.

For a positive primal candidate $z$ and dual slack
$s=b/\gamma+B^\top\eta>0$, define
\[
  \operatorname{gap}(z,s)
  :=\gamma\sum_e\bigl(s_ez_e-1-\log(s_ez_e)\bigr).
\]
In the original objective units, subtracting the dual value from the
primal objective gives
$\operatorname{gap}(z,s)+\gamma\eta^\top(b_{\rm U}-Bz)$.
Thus, when $Bz=b_{\rm U}$ holds exactly, $\operatorname{gap}(z,s)$
bounds the amount by which the primal objective exceeds its minimum.
For the raw Newton point $z_e=1/s_e$, the gap expression vanishes
regardless of the flow residual. This is why we check feasibility
separately.

We evaluate the gap using \texttt{log1p}. The stopping tolerances are
$10^{-5}$ for the scaled gap $\operatorname{gap}(z,s)/\gamma$ and
$10^{-8}$ for the maximum absolute UTC flow residual before
reconstruction. After forward flow reconstruction, we evaluate
$\operatorname{gap}(z,s)$ at the reconstructed primal point using the
dual slack from the solve. We also record the maximum absolute
residuals of the UTC flow, sequence-form flow, and fixed-point
equations.

\paragraph{Parameters and evaluation protocol.}
We fix $T$ to $1{,}000$, $100$, and $20$, and use the inverse-step parameters
\[
 \gamma_{\mathrm{K4}}=3\times10^{-5},\qquad
 \gamma_{\mathrm{K5}}=5\times10^{-6},\qquad
 \gamma_{\mathrm{Leduc}}=5\times10^{-6}.
\]
These step sizes are fixed before evaluation, and no search is performed on the reported trajectories.  The parameters are empirical choices, not the sufficient worst-case schedule in the theorem.  The dynamics, game traversal, and solvers are deterministic; all rounds are recorded, and no statistical error bars are inferred from a single deterministic run.

\paragraph{Comparators and baselines.}
The reported metric is $\max_i\LinReg_i(t)/t$, evaluated from actual played strategies.  A separate UTC backward linear-optimization recursion computes each player's best fixed linear deviation in hindsight; it does not reuse the barrier solver.  UTC-CFR/RM+ uses regret matching plus at every UTC decision node and fixed-point play \citep{zhang2024}.  Its power iteration uses a $10^{-12}$ successive-iterate tolerance and a $10{,}000$-iteration cap, and the resulting fixed-point residual is checked explicitly.

The UTC-CFR/RM+ baseline is evaluated at the same horizon as ORBITS on each game, so both methods are compared using the same full linear/UTC comparator class and the same number of rounds.


\subsection{Numerical results and verification checks}
\begin{table}[H]
\centering\footnotesize
\caption{Maximum average full linear/UTC regret at matched horizons; lower is better.}\label{tab:compact-efg}
\setlength{\tabcolsep}{4pt}
\begin{tabular}{@{}lrrr@{}}
\toprule
Game & $T$ & ORBITS & UTC-CFR/RM+\\\midrule
4-player Kuhn & 1,000 & $\mathbf{5.289\!\times\!10^{-4}}$ & $6.383\!\times\!10^{-4}$\\
5-player Kuhn & 100 & $\mathbf{3.423\!\times\!10^{-3}}$ & $4.590\!\times\!10^{-3}$\\
3-player Leduc & 20 & $\mathbf{4.357\!\times\!10^{-2}}$ & $4.758\!\times\!10^{-2}$\\
\bottomrule
\end{tabular}
\end{table}

\begin{figure}[t]
\centering
\includegraphics[width=.95\linewidth]{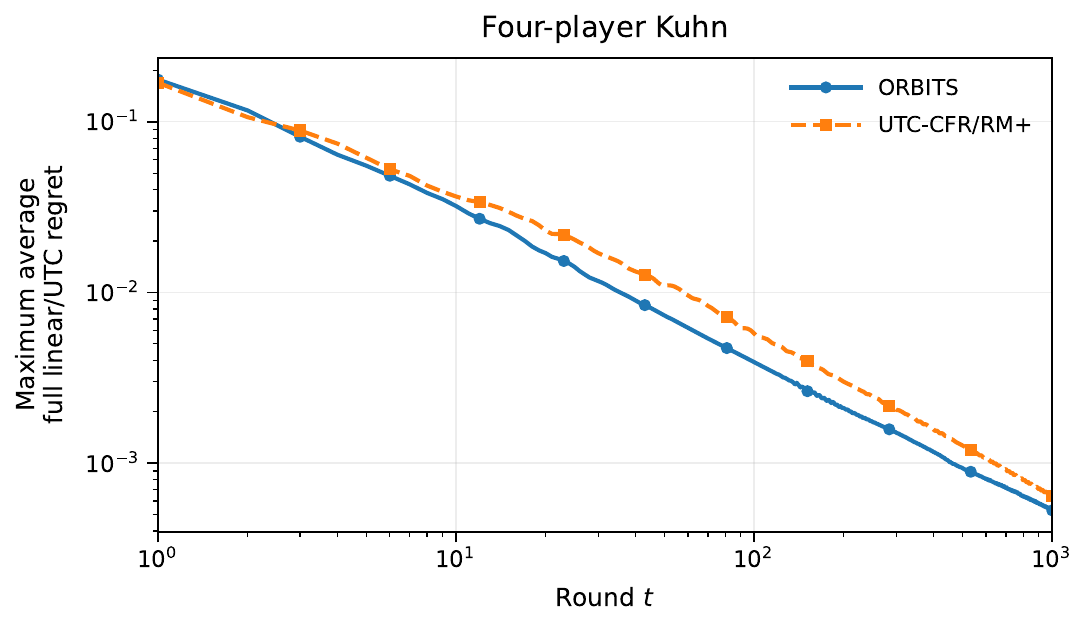}
\caption{Four-player Kuhn, all $1{,}000$ rounds.  Both axes are logarithmic; lower full linear/UTC regret is better.}
\label{fig:kuhn4-full}
\end{figure}

\begin{figure}[t]
\centering
\includegraphics[width=.95\linewidth]{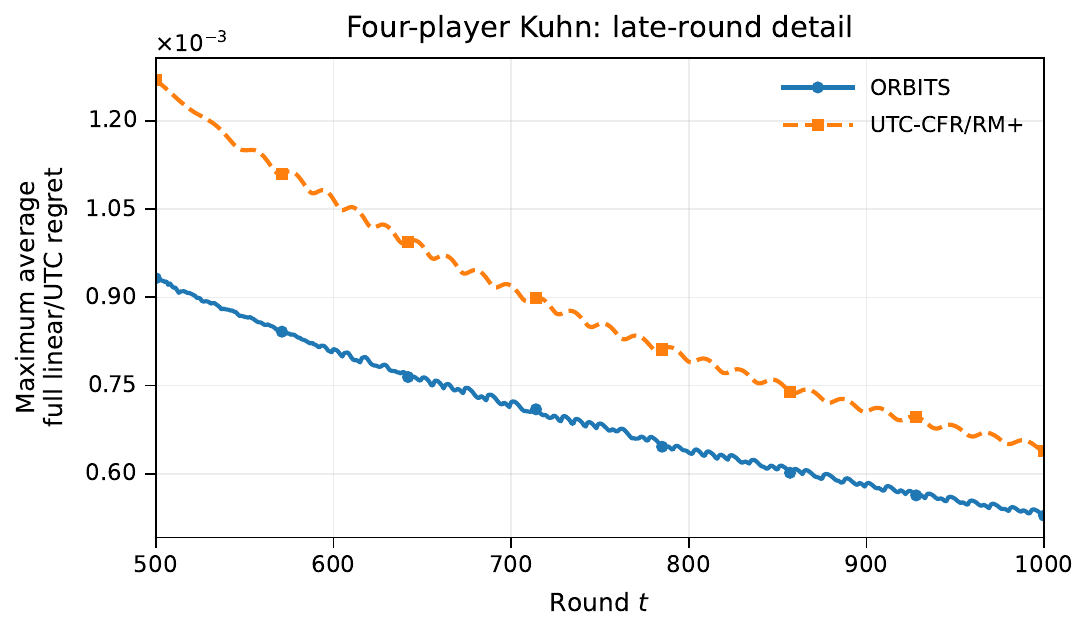}
\caption{Four-player Kuhn, rounds $500$--$1{,}000$.  Both axes are linear in this late-round detail; ORBITS remains below UTC-CFR/RM+ at the endpoint.  The full trajectory appears in Figure~\ref{fig:kuhn4-full}.}
\label{fig:kuhn4-late}
\end{figure}

\begin{figure}[t]
\centering
\includegraphics[width=.95\linewidth]{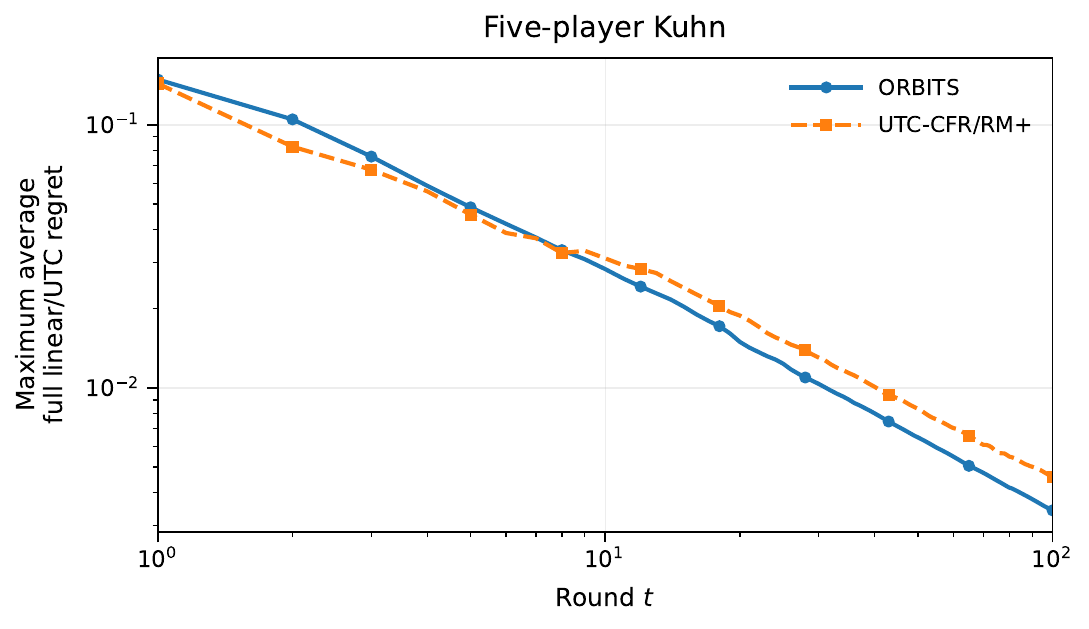}
\caption{Five-player Kuhn, all $100$ rounds.  Both axes are logarithmic; both methods are evaluated using the same full linear/UTC comparator class.}
\label{fig:kuhn5-corrected}
\end{figure}

\begin{figure}[t]
\centering
\includegraphics[width=.95\linewidth]{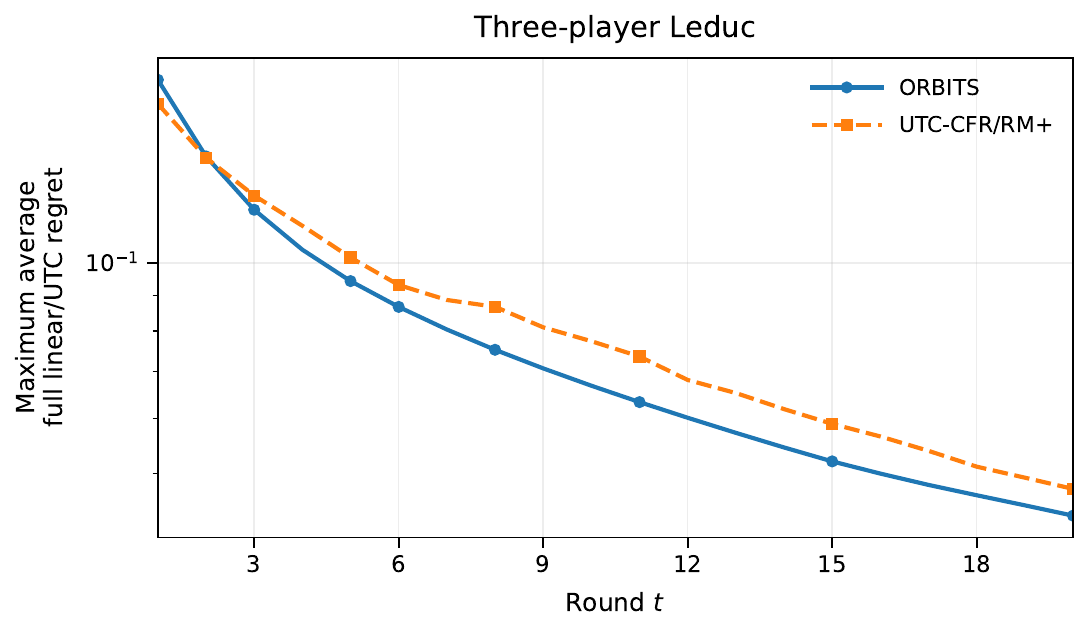}
\caption{Three-player Leduc, all $20$ rounds.  The round axis is linear and the regret axis is logarithmic.  This short-horizon comparison is not an empirical test of an asymptotic regret exponent.}
\label{fig:leduc-corrected}
\end{figure}

Table~\ref{tab:compact-efg} includes all three endpoint comparisons.  The regret curves use every recorded round without smoothing.  Figures~\ref{fig:kuhn4-full}--\ref{fig:leduc-corrected} report the complete trajectories; Figure~\ref{fig:kuhn4-late} additionally enlarges the late rounds of four-player Kuhn.

\begin{table}[H]
\centering\small
\caption{Maximum floating-point diagnostics over all ORBITS prediction
solves and players. The computed gap is $\operatorname{gap}(z,s)$
at the reconstructed primal point, in the original objective units.
The other columns give maximum absolute equation residuals.}
\label{tab:logbarrier-details}
\begin{tabular}{@{}lrrrr@{}}
\toprule
Game & Computed gap & UTC flow & Strategy flow & Fixed point\\
\midrule
4-player Kuhn & $7.51\times10^{-17}$ & $1.33\times10^{-15}$ & $1.47\times10^{-14}$ & $2.44\times10^{-15}$\\
5-player Kuhn & $3.66\times10^{-17}$ & $2.33\times10^{-15}$ & $2.59\times10^{-13}$ & $2.55\times10^{-15}$\\
3-player Leduc & $9.27\times10^{-17}$ & $4.44\times10^{-16}$ & $1.09\times10^{-14}$ & $5.11\times10^{-15}$\\
\bottomrule
\end{tabular}
\end{table}
The zero-score analytic-center solves are checked separately and saved with their residuals.  Forward reconstruction improves primal feasibility without changing the definition of the accumulated score.  

\paragraph{Independent checks.}
In the two-action example, let $u$ be the probability of querying the mediator. At the analytic center, symmetry splits the two immediate-action entries equally and does the same for each pair of actions after a recommendation. The product of the seven coordinates is proportional to $(1-u)^2u^5$, so $u=5/7$, the two immediate-action entries are $1/7$, and each of the four response entries is $5/14$. The solver recovers this center to less than $10^{-10}$ in each coordinate.  Six independent comparisons with a generic linear-program solver check the UTC hindsight oracle on the toy instance and on four-player Kuhn.  Across the three benchmarks, separate checks verify $\mathsf S\mathsf T=I$, recovery of feasible realization plans, positivity of the induced maps, and consistency of the two coordinate representations. Fixed-point residuals are recorded separately on the played profiles.  These checks are numerical implementation diagnostics; the representation and stability guarantees are proved separately in Appendix~\ref{app:utc-implementation}.

\paragraph{Tolerance sensitivity.}
The additional runs use the same step sizes and tighter tolerances of
$10^{-7}$ for the scaled gap $\operatorname{gap}(z,s)/\gamma$ and
$10^{-10}$ for the maximum absolute UTC flow residual before
reconstruction. The comparison covers the first $100$, $20$, and $3$ rounds of Kuhn-4, Kuhn-5, and Leduc, respectively.  Table~\ref{tab:tolerance-check} compares the maximum-average-regret trajectories on these prefixes.  The longer evaluation horizons were not repeated at the tighter tolerances, so this check does not certify their full-horizon numerical error.

\begin{table}[H]
\centering\small
\caption{Short-prefix tolerance checks.  The last column is the largest absolute difference between the reported maximum-average-regret curves over the indicated prefix.}
\label{tab:tolerance-check}
\begin{tabular}{@{}lrrrr@{}}
\toprule
Game & Prefix & Standard endpoint & Tighter endpoint & Max. difference\\\midrule
4-player Kuhn & 100 & $3.911\times10^{-3}$ & $3.911\times10^{-3}$ & $2.38\times10^{-8}$\\
5-player Kuhn & 20 & $1.495\times10^{-2}$ & $1.495\times10^{-2}$ & $6.46\times10^{-11}$\\
3-player Leduc & 3 & $1.192\times10^{-1}$ & $1.192\times10^{-1}$ & $3.72\times10^{-11}$\\
\bottomrule
\end{tabular}
\end{table}

\paragraph{Computing environment.}
The runs used Python 3.13.5, NumPy 2.3.5, SciPy 1.17.0, and Numba 0.65.1 on an Intel Xeon Platinum 8272CL host, with a four-core CPU quota and a 4-GiB memory limit.  Each BLAS call was limited to one thread.  ORBITS used one player worker on the two Kuhn instances and three player workers on Leduc.  Independent jobs shared the host.


